\documentclass[%
a4paper,
reprint,
superscriptaddress,
showpacs,
 amsmath,amssymb,
aps,
prb,
showkeys
]{revtex4-2}

\usepackage[utf8]{inputenc}
\usepackage{etex}
\usepackage{graphicx}
\usepackage{color}
\usepackage{colortbl}
\usepackage{xcolor}
\usepackage{calc}
\usepackage{amsfonts}
\usepackage{amsmath}
\usepackage{amssymb}
\usepackage{wasysym}
\usepackage{float}
\usepackage{epstopdf}
\usepackage{nicefrac}
\usepackage{tcolorbox}
\usepackage[makeroom]{cancel}
\usepackage{threeparttable}
\usepackage[vcentermath]{youngtab}
\usepackage{tikz}
\usepackage{rotating}
\usepackage{mathtools}
\usepackage{hyperref}
\usepackage{units}
\usepackage{tabularx}
\usepackage{pgf}
\usepackage{pgfplots}
\usepackage{geometry}
\usepackage{relsize}
\usepackage{longtable}
\usepackage{dcolumn}
\usepackage{braket}
\usepackage{listings}
\usepackage{algorithm}
\usepackage{relsize}
\usepackage{nicefrac}
\usepackage{diagbox}
\usepackage{standalone}
\usepackage{bm}
\usepackage{multirow}
\usepackage{acronym}
\usepackage{tikzorbital}
\usepackage{setspace}
\usepackage{bbold}
\usepackage[inline]{enumitem}
\usepackage{dcolumn}
\usepackage{csquotes}
\usepackage{booktabs}
\usepackage{dsfont}
\usepackage{textcomp}
\usepackage{colortbl}

\usetikzlibrary{trees}
\usetikzlibrary{calc}
\usetikzlibrary{intersections}
\usetikzlibrary{positioning}
\usetikzlibrary{shapes.misc}
\usetikzlibrary{decorations.pathreplacing}
\usetikzlibrary{shapes.arrows} 
\usetikzlibrary{backgrounds}
\usetikzlibrary{positioning,fadings,through}
\usetikzlibrary{backgrounds,fit,decorations.pathreplacing,calc}
\usetikzlibrary{decorations.pathreplacing,calligraphy}
\usetikzlibrary{arrows.meta}
\tikzset{
	solid node/.style = {circle, draw, inner sep = 3, fill = black},
	right angle/.style = {grow = 300},
	left angle/.style = {grow = 240},
	short/.style = {level distance = 1cm, },
	long/.style = {level distance = 2cm},
	left up/.style = {grow = 120},
	right up/.style = {grow = 60}
}

\pgfplotsset{compat=newest}

\newcolumntype{d}[1]{D{.}{.}{#1}}
\def \bfS  {{\bf S}}
\def \half {{\frac{1}{2}}}

\newcommand{\p}[1]{\phantom{#1}}
\newcommand{\eqnref}[1]{~(\ref{#1})}%

\providecommand{\abs}[1]{\lvert#1\rvert}

\newcommand{\der}{\partial\mspace{2mu}} 

\DeclareMathOperator{\e}{e}

\newcommand\commutator[2]{\ensuremath{\mathinner{%
			\mathopen[\,#1,#2\,\mathclose]}}}

\newcommand*{\dashfill}{\leavevmode\cleaders\hbox{-}\hfill\kern0pt}

\newcommand*{\midhrulefill}{
	\leavevmode
	\cleaders\hbox to 1ex{\raisebox{.5ex}{\rule{1ex}{.4pt}}}\hfill\kern0pt
}

\newcommand{\bracket}[2]{{\left\langle \vphantom{#1 #2} #1 \,\right|
		\left.\hspace{-0.15em} \vphantom{#1 #2} #2\right\rangle}}

\newcommand*{\braopket}[3]{\ensuremath{\langle{#1}|{#2}|{#3}\rangle}}

\renewcommand{\d}{\downarrow}
\renewcommand{\u}{\uparrow}
\newcommand{\s}{\sigma}

\newcommand{\mbf}[1]{\mathbf{#1}}

\newcolumntype{L}{>{$}l<{$}} 
\newcolumntype{C}{>{$}c<{$}} 

\newcommand{\ra}{\ensuremath{\rightarrow}}

\newcommand*{\citen}{}
\DeclareRobustCommand*{\citen}[1]{%
	\begingroup
	\romannumeral-`\x 
	\setcitestyle{numbers}%
	\cite{#1}%
	\endgroup
}

\newcommand{\cre}[1]{a_{#1}^\dagger}
\newcommand{\ann}[1]{a_{#1}^{\phantom{\dagger}}}
\newcommand{\num}[1]{n_{#1}}
\newcommand{\etal}{\emph{et al.}}
\newcommand{\Svec}[1]{\ensuremath{\hat{\mathbf{S}}_{#1}}}

\begin{document}

\author{Werner Dobrautz}
\email{dobrautz@chalmers.se}
\affiliation{%
Max Planck Institute for Solid State Research, Heisenbergstr. 1, 70569 Stuttgart, Germany
}%
\affiliation{
	Department of Chemistry and Chemical Engineering, 
	Chalmers University of Technology, 41296 Gothenburg, Sweden
}

\author{Vamshi M. Katukuri}
\affiliation{%
	Max Planck Institute for Solid State Research, Heisenbergstr. 1, 70569 Stuttgart, Germany
}%

\author{Nikolay A. Bogdanov}
\affiliation{%
	Max Planck Institute for Solid State Research, Heisenbergstr. 1, 70569 Stuttgart, Germany
}%

\author{Daniel Kats}
\affiliation{%
	Max Planck Institute for Solid State Research, Heisenbergstr. 1, 70569 Stuttgart, Germany
}%

\author{Giovanni Li Manni}
\affiliation{%
	Max Planck Institute for Solid State Research, Heisenbergstr. 1, 70569 Stuttgart, Germany
}%

\author{Ali Alavi}
\affiliation{%
Max Planck Institute for Solid State Research, Heisenbergstr. 1, 70569 Stuttgart, Germany
}%
\affiliation{
 Dept of Chemistry, University of Cambridge, Lensfield Road, Cambridge CB2 1EW, United Kingdom
}%

\title{Combined unitary and symmetric group approach applied to low-dimensional spin systems}

\date{\today}

%

\begin{abstract}
 A novel combined unitary and symmetric group approach is used to study the
spin-$\frac{1}{2}$ Heisenberg model and related Fermionic systems in a spin-adapted representation, using a
linearly-parameterised Ansatz for the many-body wave function.
We show that a more compact
ground state wave function representation is obtained when combining the symmetric group, $\mathcal{S}_n$, in
the form of permutations of the underlying lattice site ordering, with the cumulative spin-coupling
based on the unitary group, $\mathrm{U}(n)$.
In one-dimensional systems the observed compression of the wave function
is reminiscent of block-spin renormalization group approaches,
and allows us to study larger lattices (here taken up to 80 sites) with the
spin-adapted full configuration interaction quantum
Monte Carlo method, which benefits from the sparsity of
the Hamiltonian matrix and the corresponding sampled eigenstates
that emerge from the reordering.
We find that in an optimal lattice ordering the configuration state function with highest weight
already captures with high accuracy the spin-spin correlation function of the exact 
ground state wave function.
This feature is found for more general lattice models, such as the Hubbard model,
and \emph{ab initio} quantum chemical models, in this work exemplified by a
one-dimensional hydrogen chain.
We also provide numerical evidence that the optimal lattice ordering for the unitary group approach
is not generally equivalent to the optimal ordering obtained for methods based on matrix-product
states, such as the density-matrix renormalization group approach.

\end{abstract}

\maketitle

\section{Introduction \label{sec:intro}}

Symmetry is a concept of paramount importance in physics and chemistry.
Continuous symmetries are related to conservation laws by Noether's theorem\cite{Noether1918}
and are represented by Lie groups,\cite{lie-groups-general} while
discrete symmetries, given by an operator $\hat T$ that commutes with the Hamiltonian $\hat H$ of a system, are of special importance in electronic structure
calculations.
Since a set of commuting operators
can be simultaneously diagonalized, utilizing the eigenfunctions $\ket{\Phi}$ of the operator $\hat T$
causes $\hat H$ to have a block-diagonal structure in this basis.
Common discrete symmetries used in electronic structure calculations are
the discrete translational symmetry on a lattice (by the use of a momentum space basis / Bloch functions\cite{Bloch1929}), the point-group symmetries of lattices and molecules (by the use of
symmetry-adapted molecular orbitals) or conservation of the number of electron $n_\mathrm{el}$ and the
projection of the total spin $\hat S_z$ (by the use of a Slater-determinant (SD) basis with
fixed $n_\mathrm{el}$ and $m_s$).
More elaborate symmetries, such as the global $\mathrm{SU}(2)$ spin-rotation symmetry of
spin-preserving nonrelativistic Hamiltonians, necessitate a more
elaborate consideration, with the unitary group approach (UGA)\cite{Paldus1974, Paldus1976, Paldus2012, Paldus2020} being a notable example.

The fundamental postulate of quantum mechanics which states that no
observable physical quantity must change after exchanging two indistinguishable
particles leads to the concept of exchange or permutation symmetry.
The finite symmetric group $\mathcal{S}_n$ consists of all $n!$ possible permutations of $n$ objects
and, following the spin-statistics theorem,\cite{spin-statistics-theorem}
fermionic wavefunctions must transform as the antisymmetric
irreducible representation of $\mathcal{S}_n$.
Additionally, Caley's theorem states that every group and thus symmetry, can be realized as a sub-group of a symmetric group.\cite{jacobson2009basic}

In addition to the exchange of {\em particle} labels, we can also consider the effect of exchanging {\em orbital} or \emph{lattice-site} labels.
An exchange of a pair of orbitals can be seen as a 180\textdegree\, rotation between the two orbitals, a particularly simple unitary transformation of the underlying basis.\cite{feynman1999elementary}
Unlike the exchange of particle labels, which in fermionic systems leads to the aforementioned antisymmetric representation of $\mathcal{S}_n$ (and trivially realised using Slater determinants), such exchange of orbitals leads to transformations that span much larger irreducible representations of $\mathcal{S}_n$ -- in general the dimensions of these irreducible representations scale combinatorially with the number of 
orbitals in the problem. Therefore, different orderings of orbitals (or lattice sites) allow to construct different, yet equivalent Hilbert spaces.
It can be expected that, for a particular system, an optimal permutational order can be found in which the exact (e.g. ground-state) wavefunction can be expressed most compactly.

The effect of permutations of orbital/site indices can be seen as a similarity transformation of the
Hamiltonian with an
orthogonal permutation matrix, $\hat T$, connecting the ordering schemes\cite{Karwowski1997}
\begin{equation}\label{eq:sim-trans}
\e^{-\hat T} \hat H \e^{\hat T} = \bar H.
\end{equation}
There is no change of the spectrum of $\bar H$, but the explicit form of $\bar H$ in
a spin-adapted basis does change, in contrast to a SD formulation.
The influence of orbital/site ordering into the structure of the wave function 
was already observed in the application of the spin-adapted
full configuration interaction quantum Monte Carlo (GUGA-FCIQMC) 
method~\cite{Dobrautz2019, Guther2020}
to lattice systems.~\cite{dobrautz-phd}
However, the interplay between orbital type, ordering
and the unitary group, and their effects on the compactness of  many-body wave functions
was discovered in our laboratory while solving \emph{ab initio} Hamiltonians for
ground and excited states of poly-nuclear transition metal clusters,
exemplified by iron-sulfur clusters (dimers and
cubanes)~\cite{LiManni2020, LiManni2021, Dobrautz2021}, and 
manganese-oxigen tri-nuclear molecular systems.~\cite{LiManni2021b}
We found physically and chemically motivated molecular orbital unitary transformations,
based on localization and reordering,
that yield an increased compactness 
of the ground- and excited-state wave functions
(to the limit of single-reference wave functions).
Moreover, this approach leads to a unique (quasi)-block-diagonal structure of the
\emph{ab initio} Hamiltonian, that in practice allows state-specific 
optimizations of electronic excited states.
This approach is extremely beneficial for spin-adapted methods that take advantage
of the sparseness of the wave function, including GUGA-FCIQMC,
as the associated computational costs are dramatically reduced.

In the present work we study, in spin-$\frac{1}{2}$ Heisenberg systems, the
combined effect of the symmetric group $\mathcal{S}_n$ in the form of the permutations
of orbital labels and the unitary group $\mathrm{U}(n)$ providing a spin-adapted basis. Such systems exhibit large quantum fluctuations compared to aforementioned large-$S$ problems, and therefore the benefit of such transformations is not immediately obvious. Nevertheless we find that optimal orderings do exist that both compactify the exact ground-state wavefunction and in addition lead to single reference mean-field solutions whose physical properties, such as spin-spin correlation functions, are very close to the fully-correlated exact solutions. 
We will show a clear difference between the optimal orderings found 
for density matrix renormalization group (DMRG) and the one for compressing the many-body wave function within the UGA.
We find that, unlike in DMRG, it is not locality and entanglement that determine
the optimal ordering for the cumulatively spin-coupled UGA wave function, 
but a mechanism reminiscent of renormalization.
As shown in our earlier investigations and in the present work,
this finding is very general, and will be shown here for 
to Hubbard model and a chemical \emph{ab initio} model, exemplified by
a chain of hydrogen atoms.

\section{The Heisenberg model \label{sec:heisenberg}}

The Heisenberg model\cite{Heisenberg1928, Dirac1926, Dirac1929, Heisenberg1926, Vleck1934} 
describes the interaction of localized quantum-mechanical spins on a lattice and is a long-studied model, used to describe 
various aspects of magnetism in the solid state\cite{Mohn2006, Kittel1963, Nolting2009, White2007, Auerbach1994, Fazekas1999, Mattis1981, Anderson1987, Chakravarty1989, Manousakis1991, Dagotto1996, Eggert1994}.
It is given by the Hamiltonian
\begin{equation}
\label{eq:heisenberg-general}
\hat H = \sum_{ij}^n J_{ij} \Svec{i} \cdot \Svec{j}, \quad \text{with} \quad 
\Svec{i} = \{\hat S_i^x, \hat S_i^y, \hat S_i^z\}, 
\end{equation}
where the indices $i$ and $j$ run over all $n$ lattice sites, $J_{ij} = J_{ji}$ are the 
symmetric exchange constants and $\Svec{i}$ are the quantum mechanical 
spin operators with the corresponding quantum number $s \in \{\frac{1}{2}, 1, \frac{3}{2}, \dots\}$.
In this work we focus on the $s=\frac{1}{2}$ Heisenberg model with isotropic antiferromagnetic, $J_{ij} = J > 0$, nearest neighbor (NN) interactions only, indicated by the summation subscript $\braket{i,j}$ in the rest of this work.

The one-dimensional (1D) Heisenberg model with NN interaction is exactly solvable via the Bethe Ansatz\cite{Bethe1931, Karabach1997, Hulthen1938}, 
while an exact solution for higher dimensions and/or long-range interactions is still an elusive problem.\cite{Sandvik2010}
In 1D systems matrix product state (MPS) based methods\cite{Schollwoeck2011, Cirac2007, Verstraete2008, Rommer1995}, like the DMRG approach\cite{White1992, White1993, Schollwoeck20112005} are very successful, even with periodic boundary conditions\cite{Pippan2010, Dey2016} and long range interactions, due to the area law entanglement\cite{Srednicki1993, Vidal2003, Hastings2007}, while for higher dimensions 
tensor network state approaches can be applied\cite{Vidal2007, Murg2009, Ors2014, Wang2016, Li2012}.
The model does not posses a sign problem\cite{Binder1984, Loh1990, Troyer2005} for unfrustrated bipartite lattices\cite{Manousakis1991, Negele1984} and thus quantum Monte Carlo\cite{Suzuki1976, Suzuki1976a, Trotter1959,  Nyfeler2008, Sandvik2002a, Sandvik1991, Sandvik2002b, Sandvik1999, Gubernatis2016, Prokofev1998, Blankenbecler1981, Hirsch1985, Evertz1993, Evertz2003, Kalos1974, Ceperley1995, Zhang2003, Ghanem2021, Sorella1989, Hirsch1986, Assaad2008, Troyer1997, Prokofev1998, Beard1996, Foulkes2001, Tahara2008} 
and more recent neural network-based approaches\cite{Carleo2017, Melko2019, Nomura2017, Yang2020, Choo2018, Nagy2019, Choo2019, Liang2018, Chen2018, Glasser2018, Deng2017} are highly effective in providing very accurate numerical solutions 
in higher dimensions.

Due to impractical implementations, the total $\mathrm{SU}(2)$ spin symmetry of the Heisenberg model is usually not used in numerical approaches\cite{Sandvik2010, Heitmann2019}, except the work from Flocke and Karwowski\cite{Flocke1997, Karwowski1997, Karwowski1998, Karwowski2002, Flocke2002, Flocke1997b} using the symmetric group approach (SGA)\cite{sga-1, sga-2, Duch1985, Pauncz1979, Pauncz2018}, spin-symmetry adapted MPS/DMRG studies\cite{Sharma2012, McCulloch2002, Wouters2014, Xiang2001, Tatsuaki2000, Nataf2018, Singh2012, Fledderjohann2011, Singh2010} and occasional ED\cite{Bostrem2006, Heitmann2019, Bostrem2010} and real-space renormalization group studies\cite{Sinitsyn2007}.
Nevertheless the theoretical advantages of using a description conserving both total spin projection, $m_s$, the total spin, $S$, are striking:
(a) further reduction of the Hilbert space size (by additional block diagonalization of $\hat H$),
(b) optimization of electronic states of desired spin, and
(c) separation of nearly degenerate states of different total spin.

There are multiple ways to create a spin-adapted basis or the so-called configuration state functions (CSFs), see Refs.~\citen{Pauncz1979, Pauncz2018} and references therein.
One of them is the above mentioned SGA, which relies on the invariance of
the Hamiltonian with respect to permutations of electrons, or spins in the case of the Heisenberg model\cite{Flocke2002}, and it's connection to the symmetric group $\mathcal{S}_n$, being the group of all permutations of $n$ elements.
A different way to construct a spin-adapted basis is the 
UGA, pioneered by Paldus~\cite{Paldus1974, Paldus1976, Paldus2012, Paldus2020} and Shavitt~\cite{Shavitt1977, Shavitt1978, Shavitt1981, Paldus1980, Paldus1981}, 
which relies on the spin-free formulation of the  electronic structure problem~\cite{Matsen1964, Matsen1974}.
Based on Shavitt's graphical extension to UGA (GUGA),~\cite{Shavitt1977, Shavitt1978, Shavitt1981, Paldus1980, Paldus1981} 
we recently implemented a spin-adapted version of the FCIQMC method~\cite{Alavi2009, Alavi2010, Guther2020, Dobrautz2019, dobrautz-phd, Dobrautz2021}, 
which we will utilize in this work in the study of large systems beyond the capabilities of exact diagonalisation.

To some extent the influence of the ordering of orbitals in the 
GUGA was already noticed at its inception by Shavitt\cite{Shavitt1981}
and Brooks and Schaefer\cite{Brooks1979, Brooks1980}.
However, this was mostly to circumvent technical limitations of the time and did 
not concern any possible effect on the compactness of the 
ground state wavefunction. 

\section{\label{sec:theory}The spin-free Heisenberg model}

The Heisenberg Hamiltonian, see Eq.~\eqref{eq:heisenberg-general}, 
can be expressed entirely in terms of the spin-free excitation operators $\hat E_{ij} = \sum_{\sigma = \u,\d} a_{i\s}^\dagger a_{j\s}^{}$, also called shift, replacement or singlet operators\cite{Matsen1978, Duch1985, Paldus1977, Helgaker2000}, 
as follows
\begin{equation}
\label{eq:heisenberg-uga}
\hat H = -\frac{J}{2} \sum_{\braket{ij}} \hat e_{ij,ji} - \frac{J N_{b}}{4}, 
\end{equation}
where $N_b$ is the number of bonds in the lattice and $\hat e_{ij,ji} = \hat E_{ij} \hat E_{ji} - \delta_{jj}\hat E_{ii}$. 
The operator $\hat E_{ij}$ moves an electron or spin from lattice site $j$ to $i$ and fulfills the same 
commutation relations, $\commutator{\hat E_{ij}}{\hat E_{kl}} = \delta_{kj}\hat E_{il} - \delta_{il}\hat E_{kj}$, 
as the generators of the unitary group\cite{Paldus1977}. 
Flocke and Karwowski\cite{Karwowski1997}, 
employed the related SGA to study the Heisenberg model in a 
spin-adapted way.
The analogy between the SGA and UGA formulation of the Heisenberg Hamiltonian is reviewed in Appendix~\ref{app:dirac}.
Eq.~\eqref{eq:heisenberg-uga} allows us to study the Heisenberg model in a spin-adapted formalism via the UGA, as utilized in the GUGA-FCIQMC method.

\section{Spin eigenfunctions and the action of permutation operators}
Given a system of $n$ lattice sites, each associated with a spin-$\frac{1}{2}$ variable,
$\sigma_i=\u,\d$, the Hilbert space of $2^n$ primitive spin functions $\ket{\sigma_1\sigma_2...\sigma_n}$
can be subdivided into $\binom{n}{(n+m_s)/2}$ states with a fixed spin
polarization $m_s$.
From this set $g(n,S)=\binom{n}{n/2-S}- \binom{n}{n/2-S-1}$ spin eigenfunctions\cite{Pauncz1979}
with spin $S$ and $m_s = S$ can be constructed.
The set forms a $g$-dimensional {\em irreducible} representation
of the permutational group.\cite{Wigner1931}
In this section we outline, through some simple examples, the consequences of this
fact, which we will later exploit
more generally.

Spin eigenfunctions can be constructed {\em geneologically} \cite{Salmon1974, Pauncz1979} using the addition theorem of angular momentum. Thus an $n$-electron CSF with spin $S$ can be constructed from an
$(n-1)$ electron CSF with spin $S\pm \half$, by positively or negatively spin-coupling with a spin $s=\half$. A positive spin coupling $\Delta S=+\half$ is denoted with a symbol $u$ (up-spin) and a negative $\Delta S=-\half$ with $d$ (down-spin).
In the GUGA method, this construction is carried out cumulatively, starting with a single spin, and adding one spin at a time, until the $n$-electron CSF is constructed. An $n$-electron CSF is therefore denoted as a string of $n$ $u$'s and $d$'s, such as $\ket{uudd}$.
At each intermediate step, say step $i$, a pure-spin CSF is obtained, with cumulative spin $S_i=\sum_j^i \Delta S_j \ge 0$. This means that first element of the CSF string must be a $u$, and at each step of this cumulative construction, the  number of $d$'s cannot exceed that of $u$'s. Also, the final cumulative spin $S_n=S$.
For example, for an $n=3$ system with $S=\frac{1}{2}$, there are $g=2$ CSFs ($\ket{uud}$ and $\ket{udu}$) which can be constructed, from a 2-electron triplet and singlet, respectively:
\begin{eqnarray}
\ket{u u d} & =& \frac{1}{\sqrt{6}}\big(2 \ket{\u \u \d} - \ket{(\u \d+\d\u)\u}  \big )  \\
\ket{u d u} & =& \frac{1}{\sqrt{2}}\ket{(\u\d-\d\u)\u}
\end{eqnarray}
Note that, in the above, we have not explicitly labelled the spins:
we might assume them to be in natural order $1,2,...,n$.
However, this is not necessary, and we are at liberty to construct the CSFs by coupling 
the spins in any order we like.
To be specific, let us insert the labels of the sites, for example in natural order:
\begin{eqnarray}
\ket{u_1 u_2 d_3} & =& \frac{1}{\sqrt{6}}\big(2 \ket{\u_1 \u_2 \d_3} - \ket{(\u_1 \d_2+\d_1\u_2)\u_3}  \big )  \\
\ket{u_1 d_2 u_3} & =& \frac{1}{\sqrt{2}}\ket{(\u_1 \d_2-\d_1\u_2)\u_3}
\end{eqnarray}
Consider the action of the permutation operator $P_{23}$:
\begin{eqnarray}
P_{23}\ket{u_1u_2d_3} & = & \ket{u_1 u_3 d_2} \nonumber \\
& = & \frac{1}{\sqrt{6}}\big(2 \ket{\u_1 \u_3 \d_2} - \ket{(\u_1 \d_3+\d_1\u_3)\u_2}  \big ) \nonumber \\
& = & -\frac{1}{2} \ket{u_1 u_2 d_3} + \frac{\sqrt{3}}{2} \ket{u_1 d_2 u_3} \\
P_{23} \ket{u_1d_2u_3} & = & \ket{u_1 d_3 u_2} \\
&= & \frac{\sqrt{3}}{2}\ket{u_1 u_2 d_3}  + \frac{1}{2} \ket{u_1 d_2 u_3}
\end{eqnarray}
In this way, we generate an {\em equivalent} set of spin eigenfunctions. However,
if we consider the Heisenberg Hamiltonian with $J=1$ for the 3-site chain with
open boundary conditions:
\begin{equation}
\hat H=  \hat{\bfS}_1 \cdot \hat{\bfS}_2 + \hat{\bfS}_2 \cdot \hat{\bfS}_3.
\end{equation}
In the natural  order (1-2-3), the $S=\half$ sector of the Hamiltonian is represented as
\begin{equation}\label{eq:csf-3-site}
H^{123} = -
\begin{pmatrix}
\frac{5}{4} & \frac{\sqrt{3}}{4} \\
\frac{\sqrt{3}}{4} & \frac{3}{4} \\
\end{pmatrix},
\end{equation}
while in the second order (1-3-2), it is {\em already} diagonal:
\begin{equation}\label{eq:csf-3-site-compact}
H^{132} =
\begin{pmatrix}
-\frac{3}{2} & 0 \\
0 & -\frac{1}{2} \\
\end{pmatrix}.
\end{equation}
Therefore, through a process of mere site re-ordering in the CSFs, we have diagonalized the Hamiltonian,
meaning that a single CSF is able to fully capture the exact eigenstates of the Hamiltonian in this
spin sector of the Hamiltonian.
As the two representations of $\hat H$ in Eq.~\eqref{eq:csf-3-site} and Eq.~\eqref{eq:csf-3-site-compact}
are related by a similarity transformation generated by the
orthogonal permutation matrix connecting the ordering schemes~\cite{Karwowski1997}, 
see Eq.~\eqref{eq:sim-trans}, the eigenvalues do not change, while the eigenvectors undergo a very advantageous compression.

Moving to the 4-site Heisenberg chain with periodic boundary conditions, we find a similar behavior. Here the Hamiltonian is given by:
\begin{equation}
\hat H=  \hat{\bfS}_1 \cdot \hat{\bfS}_2 + \hat{\bfS}_2 \cdot \hat{\bfS}_3+ \hat{\bfS}_3 \cdot \hat{\bfS}_4+ \hat{\bfS}_4 \cdot \hat{\bfS}_1
\end{equation}
The $S =0$ CSF Hilbert space is still only 2 dimensional ($g(4,0)=2$), with the two states $\ket{uudd}$ and $\ket{udud}$.
Employing the natural order 1-2-3-4, we obtain the Hamiltonian
\begin{equation}\label{eq:csf-4-site-orig}
H^{1234} = -
\begin{pmatrix}
\frac{1}{2} & \frac{\sqrt{3}}{2} \\
\frac{\sqrt{3}}{2} & \frac{3}{2}
\end{pmatrix}
\end{equation}
whilst the order 1-3-2-4 yields the diagonal Hamiltonian
\begin{equation}\label{eq:csf-4-site-compact}
H^{1324} = -
\begin{pmatrix}
2 & 0 \\
0 & 0
\end{pmatrix},
\end{equation}
implying once again single-CSF exact eigenstates.
This procedure is very general and of broad applicability, 
as we have shown in Refs.~\citen{LiManni2020, LiManni2021} 
for the general non-relativistic \emph{ab initio} molecular Hamiltonian 
in the case of a nitrogen dimer and an iron-sulfur cubane chemical model system.

It is worth noting that the 4-site chain with {\em open} boundary conditions is not brought into diagonal form by any ordering.
No low-spin ground state of a Heisenberg chain/ring of size $n > 4$ with NN interaction only is exactly \enquote{single reference}.
Nevertheless, as we shall show below, the underlying site ordering largely influences the weight and nature of
the most dominant CSF (called the \enquote{reference CSF}).
It is precisely this interplay between permutations of site indices, represented by the symmetric group $\mathcal{S}_n$, and the spin-adapted basis given by the unitary group approach, $\mathrm{U}(n)$, which we wish to study in this paper.

We also note that Flocke and Karwowski\cite{Karwowski1997} briefly mention the effect of the
numbering of the lattice sites on the SGA in their 1997 paper. They found that the ordering affects the
number of necessary matrix multiplications to construct a transposition $(i,j)$.
They illustrated the effect on a $4\times 2$ square lattice (ladder),
and although we focus on 1D systems in this work, we checked to
find that they find a different ordering compared to our case, since they were motivated
to minimize the number of matrix multiplications to construct the transposition $(i,j)$,
see the Supplemental Material~\cite{SI}, whereas our motivation is to find representations in which the exact
wavefunctions assume maximally compact forms.

\subsection{\label{sec:motivation}Extension to larger systems}

Due to the factorial growth of the number of possible reorderings with the system size, it
is important to have a physically-motivated approach for the relabeling for larger lattices.
An obvious choice for a 6-site chain with open boundary conditions (OBC),
or ring with periodic boundary conditions (PBC), would be a \enquote{bipartite} ordering,
with one possible bipartite ordering given by e.g. (1-3-5-2-4-6).
Indeed, such a reordering, increases the weight of the most dominant CSF,
given by $\ket{u_1 u_3 u_5 d_2 d_4 d_6}$ to 95.7\% compared to the natural order
reference state $\ket{u_1 d_2 u_3 d_4 u_5 d_6}$ with a weight of 77.9\% for PBC.
However, for OBC, this bipartite ordering actually decreases the weight of the leading CSF to 89.9\% compared
with 92.2\% for the natural ordering (see Table~\ref{tab:6-site-orders}).

\begin{table}
	\caption{\label{tab:6-site-orders}6-site orderings and leading CSF weights.}
	\renewcommand{\arraystretch}{1.1}
	{\small
		\begin{tabular}{ccccc}
			\toprule
			& & & \multicolumn{2}{c}{CI coefficient [\%]} \\
			\multicolumn{2}{c}{Order} & Ref. CSF & PBC & OBC \\
			\midrule
			Natural   & 1-2-3-4-5-6 & $\ket{u_1 d_2 u_3 d_4 u_5 d_6}$ & 	77.9 & 92.2 \\
			Bipartite & 1-3-5-2-4-6 & $\ket{u_1 u_3 u_5 d_2 d_4 d_6}$ & 	95.7 & 89.9	\\
			Compact   & 1-3-2-5-4-6 & $\ket{u_1 u_3 d_2 u_5 d_4 d_6}$ & 	97.1 & 94.7 \\
			\midrule
			\multicolumn{2}{c}{SDs Néel state} & $\ket{\u_1 \d_2 \u_3 \d_4 \u_5 \d_6}$ & 47.9 & 44.9 \\
			\bottomrule
		\end{tabular}
	}
\end{table}

\begin{figure}
	\includegraphics[width=0.45\textwidth]{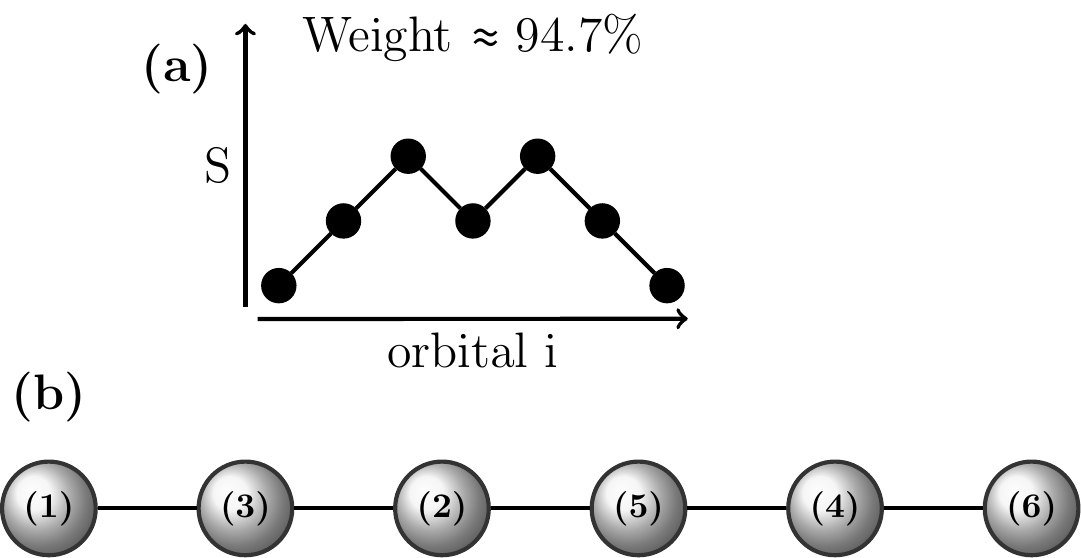}
	\caption{\label{fig:6-site-open}Weight of leading CSF for the compact ordered 6 site chain with OBC. The labels on the lattice sites refer to the order in which the spins are coupled in the GUGA CSF formalism. Thus, for example, the second site from left with label \enquote{3} is coupled in the 3rd position in the CSF.}
\end{figure}

\begin{figure}
	\includegraphics[width=0.45\textwidth]{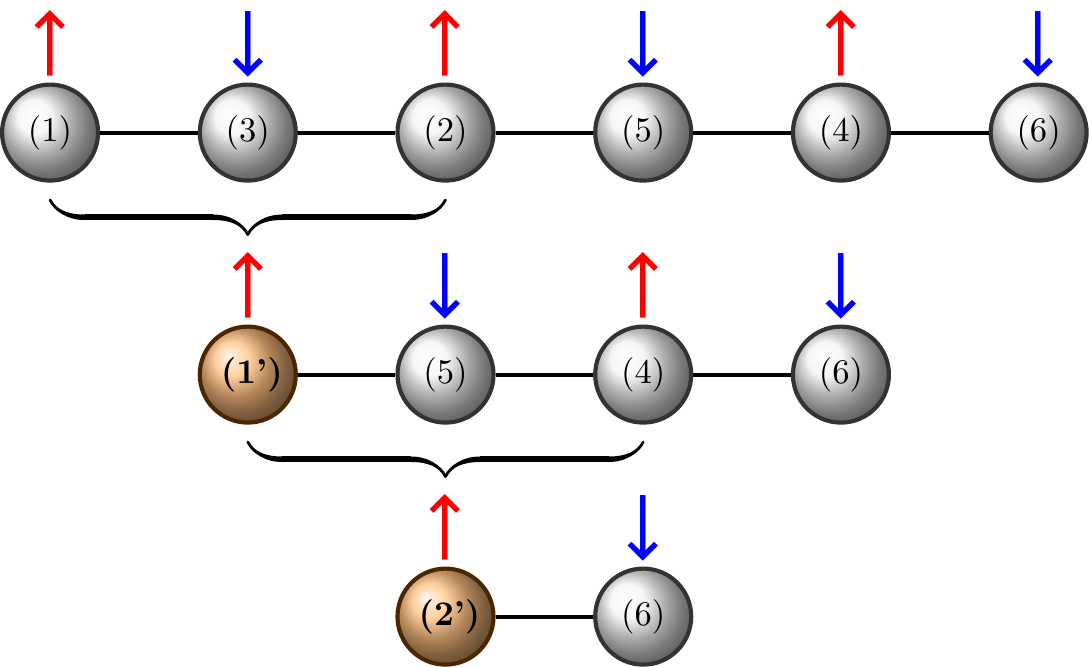}
	\caption{\label{fig:6-site-renorm}Cumulative doublet coupling of \enquote{meta-spin-$\frac{1}{2}$} in the most compact order.}
\end{figure}

Very interestingly, we find that there exists an even \enquote{more optimal} ordering, which we term \emph{compact} ordering, shown in Fig.~\ref{fig:6-site-open}, to be explained in the following:
Let us take a closer look at the (1-3-2) and (1-3-2-4) orderings of the 3- and 4-site lattice.
The (1-3-2) ordering in the 3-site OBC case leads to a single-CSF doublet ground state,
and this fact is responsible for the massively increased weight of the leading CSF also in larger 1D Heisenberg systems.
As we are dealing with a $S=\frac{1}{2}$ Heisenberg model, each physical site is locally a doublet.
Similar to renormalization group approaches\cite{Fisher1974, Hoffmann2020, Malrieu2001, Wilson1975}, one can interpret that the first three sites under the ordering (1-3-2) are coupled to a doublet state $S=\frac{1}{2}$ with the CSF $\ket{uud}$, termed \enquote{three-site meta-spins $\frac{1}{2}$} by Malrieu\cite{Hoffmann2020}, and reminiscent of the block spin idea by Kadanoff\cite{Kadanoff1966} (see Figure~\ref{fig:6-site-renorm}).
To confirm this, we measured the \emph{local spin} expectation value of the first three sites (described in the Appendix of Ref.~\citen{Dobrautz2021})and obtained a value very close the expected value of a doublet ($\frac{3}{4}$), $\braket{(\hat{\mathbf{S}}_1 + \hat{\mathbf{S}}_3 + \hat{\mathbf{S}}_2 )^2} \approx 0.751$.

Thus, if we interpret sites (1-3-2) now as a renormalized site \textbf{1'}, we can again couple three \enquote{sites}, (\textbf{1'}-5-4)
to a new doublet with index \textbf{2'}, again with $\ket{\mbf{u\text{\bf'}}ud}$, where $\mbf{u\text{\bf'}}$ indicates the renormalized doublet.
Finally, for the 6-site system the renormalized doublet couples to singlet with the remaining 6th site, yielding the total reference CSF as $\ket{u_1 u_3 d_2 u_5 d_4 d_6}$. This process is schematically displayed in Fig.~\ref{fig:6-site-renorm} and the corresponding site ordering and genealogical spin-coupling of the 
leading CSF in the compact ordering are shown in Fig.~\ref{fig:6-site-open}.
The weight of this CSF in the compact ordering is 97.1\% for PBC and 94.7\% for OBC in the 6-site lattice,
a much larger weight as compared to both the natural and bipartite orderings (see Table~\ref{tab:6-site-orders}).

One important difference emerges from the comparison of the natural ordering and the compact renormalized ordering.
In the normal ordering the leading CSF ($\ket{ududud}$) is such that at every second site the cumulative spin vanishes ($S_i=0$). 
The long-range spin correlation is therefore transferred to the other CSFs of the ground state wave function.
In the compact renormalized ordering at no time $S=0$ in the leading CSF (except for the last site for a total $S=0$ state with an even number of sites).
We describe this feature as a \textit{propagating doublet} along the chain.
Thus, already the leading CSF carries information on the long range correlation. This aspect will be discussed further in the following.

An expression for a total $n$-electron singlet CSF $\ket{u(ud)^{\frac{n-2} {2}}d}$, arising from the coupling
of the propagating doublet with the last spin, reads in second quantized form
\begin{equation}
\label{eq:recurr_sing}
\Psi_{n}^{S=0}=\frac{1}{\sqrt{2}} \left [ \psi_{(n-1) \u} a^\dagger_{n \d} - \psi_{(n-1) \d} a^\dagger_{n \u} \right ],
\end{equation}
where $a^\dagger_{i \sigma}$ is a creation operator at (ordered) position $i$ with spin $\sigma\!\in\!\left \{ \u, \d \right \}$ and $\psi_{i \sigma}$ is the propagating
doublet defined by a recurrent formula
\begin{equation}
\begin{aligned}
\label{eq:recurr_doub}
\psi_{i \sigma} &= C^{uud}_\sigma \psi_{(i-2) \sigma}a^\dagger_{(i-1) \sigma} a^\dagger_{i \bar{\sigma}}  \\
&+ C^{udu}_\sigma \left ( \psi_{(i-2) \sigma} a^\dagger_{(i-1) \bar{\sigma}}+ \psi_{(i-2) \bar{\sigma}} a^\dagger_{(i-1) \sigma} \right ) a^\dagger_{i \sigma}
\end{aligned}
\end{equation}
with $i\!\in\!\left \{ 1, 3, 5, ..., n-1 \right \}$ and base $\psi_{1 \sigma}=a^\dagger_{1\sigma}$.
The numerical coefficients $C^{uud}_\sigma$ and $C^{udu}_\sigma$ are defined by the Clebsch-Gordan coefficients
$\bracket{s_1 m_1; s_2 m_2}{s_\mathrm{tot} m_\mathrm{tot}} $
arising from the angular momenta addition as
\begin{equation}
\begin{aligned}
&C^{uud}_\u = \bracket{\frac{1}{2} \frac{1}{2}; \frac{1}{2} \frac{1}{2}}{1 1}
\bracket{1 1; \frac{1}{2} \bar{\frac{1}{2}}}{\frac{1}{2} \frac{1}{2}} = \sqrt{\frac{2}{3}}, \\
&C^{uud}_\d = \bracket{\frac{1}{2} \bar{\frac{1}{2}}; \frac{1}{2} \bar{\frac{1}{2}}}{1 \bar{1}}
\bracket{1 \bar{1}; \frac{1}{2} \frac{1}{2}}{\frac{1}{2} \bar{\frac{1}{2}}} = -\sqrt{\frac{2}{3}} \\
\end{aligned}
\end{equation}
and
\begin{equation}
\begin{aligned}
&C^{udu}_\u = \bracket{\frac{1}{2} \frac{1}{2}; \frac{1}{2} \bar{\frac{1}{2}}}{1 0}
\bracket{1 0; \frac{1}{2} \frac{1}{2}}{\frac{1}{2} \frac{1}{2}} = -\sqrt{\frac{1}{6}}, \\
&C^{udu}_\d = \bracket{\frac{1}{2} \bar{\frac{1}{2}}; \frac{1}{2} \frac{1}{2}}{1 0}
\bracket{1 0; \frac{1}{2} \bar{\frac{1}{2}}}{\frac{1}{2} \bar{\frac{1}{2}}} = \sqrt{\frac{1}{6}}\,.
\end{aligned}
\end{equation}
One could also think about separate coupling of sites (1-3-2) and (6-5-4) to two doublet states,
and a consequent coupling of the two doublet states to an overall singlet,
as described in Ref.~\cite{Hoffmann2020}.
Interestingly, for the 6-site ring, both these approaches are identical and lead exactly to the same Hamiltonian representation.
However, as our ultimate goal is to study these systems with GUGA-FCIQMC method,
we focus on the cumulative approach here.

With Eq. \eqref{eq:recurr_sing} and\eqref{eq:recurr_doub}, the described ordering is easy to generalize to larger lattice sites.
Moreover, it is not restricted to a bipartite lattice with an even number of sites, but is also applicable to inherently
frustrated systems with an odd number of sites, as shown for the weights of the leading CSFs for a 7-site lattice with PBCs in
Table~\ref{tab:7-site-orders}.

\begin{table}
	\caption{\label{tab:7-site-orders}7-site orderings and leading CSF weights.}
	\renewcommand{\arraystretch}{1.1}
	{\small
		\begin{tabular}{ccccc}
			\toprule
			& & & \multicolumn{2}{c}{CI coefficient [\%]} \\
			\multicolumn{2}{c}{Order} & Ref. CSF & PBC & OBC \\
			\midrule
			Natural   & 1-2-3-4-5-6-7 & $\ket{u_1 d_2 u_3 d_4 u_5 d_6 u_7}$ & 	79.5 & 67.0 \\
			Compact   & 1-3-2-5-4-7-6 & $\ket{u_1 u_3 d_2 u_5 d_4 u_7 d_6}$ & 	88.7 & 94.3 \\
			\midrule
			\multicolumn{2}{c}{SDs \enquote{N\'{e}el state}} & $\ket{\u_1 \d_2 \u_3 \d_4 \u_5 \u_6 \d_7}$ & 40.5 & 54.4 \\
			\bottomrule
		\end{tabular}
	}
\end{table}

\subsection{\label{sec:brute-force}Exhaustive search study}

We confirmed our renormalization-group motivated\cite{Hoffmann2020, Malrieu2001} Ansatz, described above, by
considering all $n!$ possible permutations of site labels.
For all these permutations we exactly diagonalized the corresponding Heisenberg Hamiltonian and investigated what the highest weighted CSF in the ground state is.
Due to the rather small Hilbert space size, we were able to exhaustively  search the full $n$-factorial permutational space
up to 10 site systems and confirm that this compact ordering holds for these system sizes.
The optimal ordering for a 10 site system is (1-3-2-5-4-7-6-9-8-10)
and the dominant CSF is $\ket{u_1 u_3 d_2 u_5 d_4 u_7 d_6 u_9 d_8 d_{10}}$ with a 90.3\% weight,
which again reflects the coupling of three consecutive sites to a doublet
in an iterative way, as depicted in Fig.~\ref{fig:6-site-renorm}.

\begin{figure}
	\includegraphics[width=0.4\textwidth]{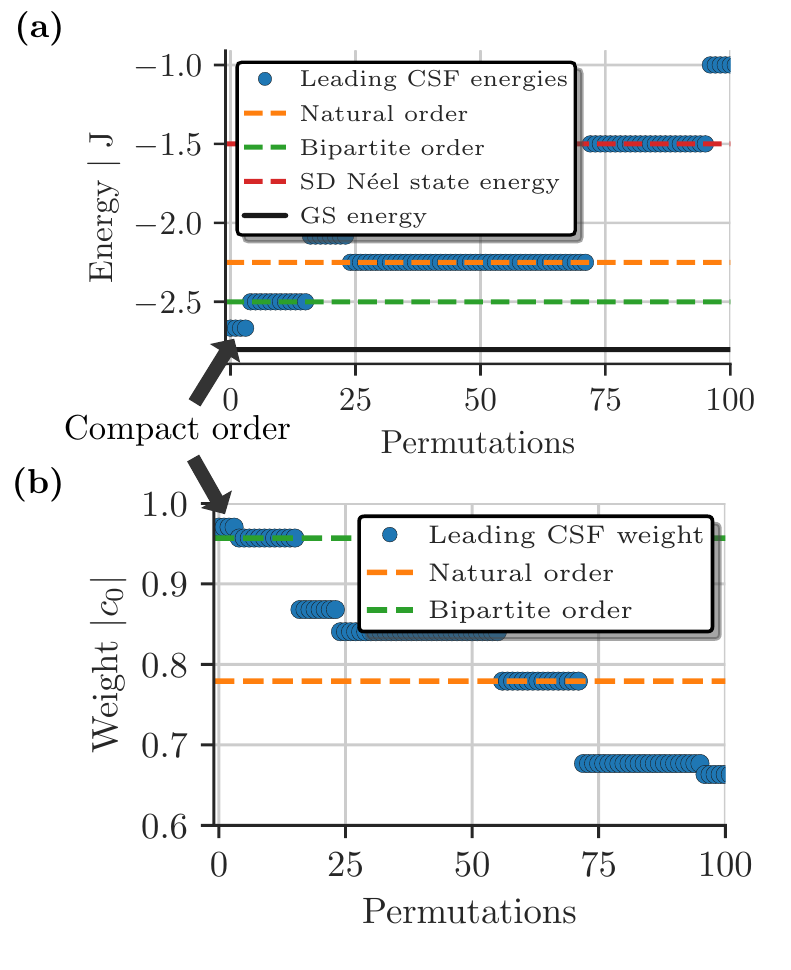}
	\caption{\label{fig:6-site-ring-ref-e}(a) Lowest single CSF energies
		and (b) corresponding weight $\abs{c_0}$  for the 6-site ring as a function of permutations.
	}
\end{figure}

For systems larger than 10 sites the combinatorial growth of permutations ($10! = 3628800$ already)
prevents an exhaustive search.
At the same time obtaining the highest weighted CSF requires a diagonalization of the Hamiltonian, which
also is not a feasible route to scale to bigger problems.
For this reason we investigated possible cheaper indicators of an optimal ordering and the corresponding leading CSF for system sizes up to 10 sites.
For the spin-$\frac{1}{2}$ Heisenberg model with nearest neighbor interaction the diagonal matrix elements are an indicator for the optimal ordering.
Fig.~\ref{fig:6-site-ring-ref-e} shows the diagonal matrix elements of the highest-weighted CSF as a function
of the possible non-cyclic permutations (only the first 100 are shown for clarity) of a 6-site ring.
The first 4 data points correspond to 4 equivalent compact orderings, followed by 12 equivalent bipartite orderings of the site labels.
The figure also shows the single CSF energies of the natural ordering and the diagonal element of a N\'{e}el state in a SD basis. One can see there is a stark decrease of the diagonal matrix elements in the compact ordering, which
is already quite close to the ground state energy of this system.
Thus, a cheaper option to find the most compact representation, other than diagonalization of the full Hamiltonian, is
to minimize the diagonal matrix element over the space of possible permutations
\begin{equation}
\min_{\mathcal{S}_n} \min_{\mu}\, \braopket{\mu}{H}{\mu}.
\end{equation}
In the GUGA-Heisenberg model the single CSF energies are given by the diagonal exchange contributions\cite{Dobrautz2019, Shavitt1978, dobrautz-phd}:
\begin{equation}\label{eq:diagonal-exchange}
\braopket{\mu}{\hat H}{\mu} \sim \frac{1}{2}\sum_{j > i} J_{ij} X_{ij}(\mu), \quad J_{ij} =\begin{cases}
J & \text{for NN},\\
0 & \text{else},
\end{cases}
\end{equation}
where $X_{ij}(\mu)$ is a CSF dependent quantity and $J_{ij}$
is non-zero for NN sites $i$ and $j$ depending on the chosen ordering.
For NN interaction only, the task to minimize Eq.~\eqref{eq:diagonal-exchange} for a given
CSF $\ket{\mu}$ is equivalent to the traveling salesman problem (TSP)\cite{Applegate2006} (after scaling the possibly negative $X_{ij}(\mu)$ to positive quantities).
This can best be seen when formulating the TSP as a integer linear program\cite{Papadimitriou1998, Dantzig1963} in the Miller-Tucker-Zemlin\cite{Miller1960} or Dantzig-Fulkerson-Johnson formulation\cite{Dantzig1954}.
For more general Heisenberg (higher dimension, longer and anisotropic interactions), or even \emph{ab initio} models, the minimization of the corresponding Eq.~\eqref{eq:diagonal-exchange} can be mapped to more general quadratic assignment problem\cite{Koopmans1957}.

\begin{figure*}
	\includegraphics[width=0.8\textwidth]{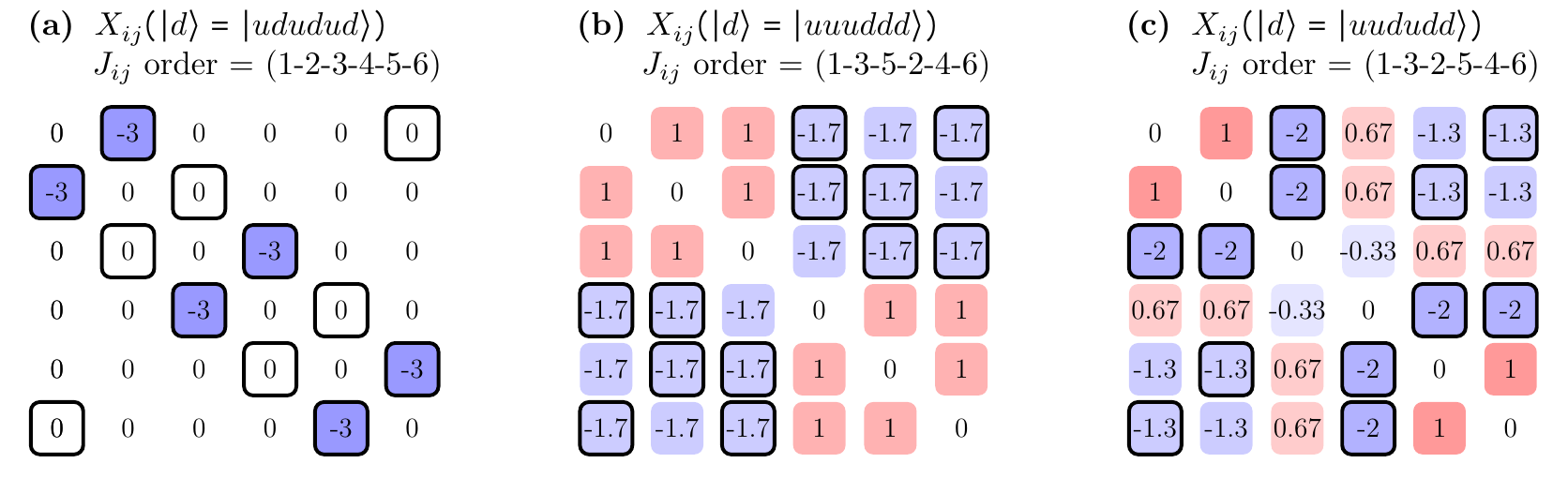}
	\caption{\label{fig:6-site-exchange-orig} Exchange matrix elements, $X_{ij}$, for the 6 site chain 
		reference CSFs: (a) $\ket{ududud}$ (natural), (b) $\ket{uuuddd}$ (bipartite) and (c) $\ket{uududd}$ (compact), with the respective order in parenthesis. 
		The non-zero $J_{ij}$ values, due to the corresponding orderings are indicated by the black rectangles.
		They act as a \enquote{mask} in the product $X_{ij} J_{ij}$, see Eq.~\eqref{eq:diagonal-exchange}, 
		and for a given CSF, the \enquote{optimal} lattice ordering yields the lowest possible diagonal matrix element, Eq.~\eqref{eq:diagonal-exchange}. 
	}
\end{figure*}

Both quantities $X_{ij}(\mu)$ and $J_{ij}$ can be expressed in matrix form, where the task is to construct an
ordering of the 1D system, which changes $J_{ij}$ to give the lowest possible diagonal matrix element, according to Eq.~\eqref{eq:diagonal-exchange}.
Fig.~\ref{fig:6-site-exchange-orig} shows $X_{ij}(\mu)$ for (a) $\ket{ududud}$ with the original/natural ordering (1-2-3-4-5-6), (b) $\ket{uuuddd}$ with the bipartite
ordering (1-4-2-5-3-6) and (c) $\ket{uududd}$ with the \emph{compact} ordering (1-3-2-5-4-6). 
The optimal $J_{ij}$ and thus orderings, which minimize the diagonal matrix element, Eq.~\eqref{eq:diagonal-exchange}, are indicated by the red squares, which act as a \enquote{mask} and 
determine which $X_{ij}$ elements contribute to the sum in Eq.~\eqref{eq:diagonal-exchange} (indicated by a red background color in Fig.~\ref{fig:6-site-exchange-orig}).
The direct relation between the lowest diagonal matrix element corresponding to the highest weight in the ground state wavefunction can be used to implement efficient approximate solvers to find the optimal permutation and
confirm our assumed renormalization structure.
We implemented a simulated annealing\cite{Khachaturyan1981, vanLaarhoven1987, Kirkpatrick1983, Cern1985, Pincus1970} minimizer with 2-\cite{Croes1958, Flood1956} and 3-opt modifications\cite{Lin1965} based on the Lin-Kernigham heuristic\cite{Lin1973} to find the optimal ordering for a given CSF.
Additionally, the mapping to the TSP allowed us to find the optimal ordering with
a state-of-the-art solver by Helsgaun\cite{Helsgaun2000, Helsgaun2009} 
(see the Supplemental Material for sample input files\cite{SI}). 

To deal with the exponentially growing Hilbert space with increasing system size
we combined our excitation generation routines of FCIQMC to stochastically suggest new states, $\ket{\nu}$
for a given CSF, $\ket{\mu}$.
With this approach we were able to confirm our renormalized ordering Ansatz for system sizes
up to $n = 20$ (where we were still able to enumerate the whole Hilbert space),
which gives us great confidence that this is not a result restricted to small system sizes.
More details on our simulated annealing approach can be found in Appendix~\ref{sec:annealing}.

\subsection{\label{sec:spin-corr}Spin-spin correlation functions}

\begin{figure*}
	\includegraphics[width=\textwidth]{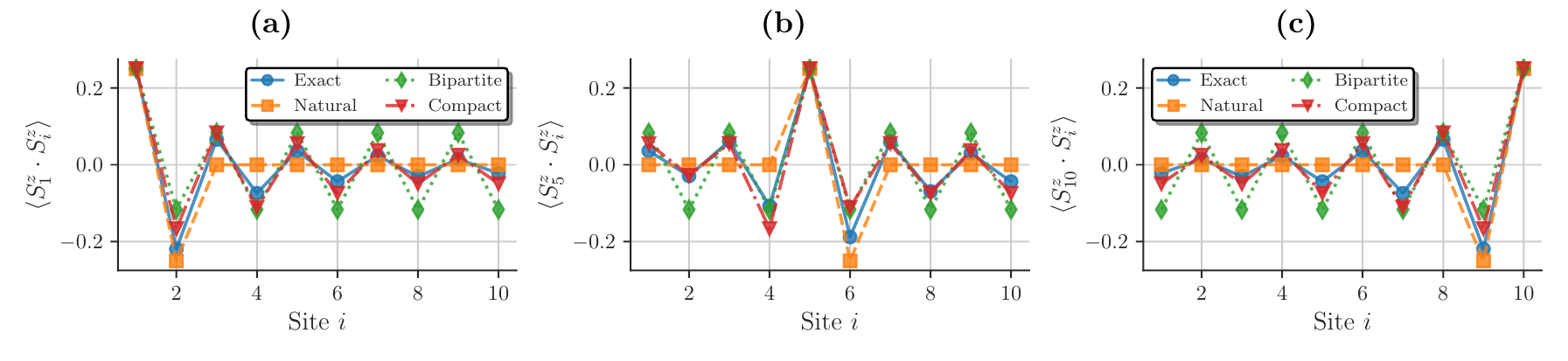}
	\caption{\label{fig:10-site-single-spin-corr} (a) $\braket{S_1^z \cdot S_x^z}$, (b) $\braket{S_5^z \cdot S_x^z}$ and (c) $\braket{S_{10}^z \cdot S_x^z}$ exact and single-CSF spin-spin correlation functions for the 10 site chain with OBC.}
\end{figure*}

In addition to the renormalization behavior,
another physical motivation can be drawn for
the dominant CSF in the optimal ordering:
it shows a spin-spin correlation function similar to the exact ground state wave function.
This is possible, since a CSF, $\ket{\mu}$, consists of a linear combination of SDs\cite{Shavitt1981, Harter1976, Olsen2014, Fales2020}, $\{\ket{I}\}$, (see Supplemental Material~\cite{SI} for examples)
\begin{equation}
\label{eq:lincomb}
\ket{\mu} = \sum_I c_I \ket{I},
\end{equation}
and thus can yield non-trivial spin-spin correlation functions, even for a single CSF.
Fig.~\ref{fig:10-site-single-spin-corr} shows three spin-spin correlation functions $\braket{\hat S_j^z \cdot \hat S_x^z}$,
with $j = 1, 5$ and $j=10$ for a 10 site Heisenberg chain with OBC,
for the different lattice orderings considered in this work compared with the exact result.
Both the natural and bipartite single CSF spin-correlation function show a rather trivial behavior. The former characterized by a quickly vanishing
spin-spin correlation function already after the first NN, and the latter exhibiting a 
alternating uniform 
correlation function. 
In the case of the natural order leading CSF, $\ket{u_1 d_2 u_3 d_4 u_5 d_6 u_7 d_8 u_9 d_{10}}$, 
this behavior stems from the above mentioned vanishing intermediate spin, $S_i = 0$, at every other lattice site. 
On the contrary, already the leading CSF in the \textit{compact} ordering 
carries information of the long-range spin-correlation and
exhibits a spin-correlation
function that resembles the exact $\braket{\hat S_j^z \cdot \hat S_x^z}$.

\begin{figure*}
	\includegraphics[width=\textwidth]{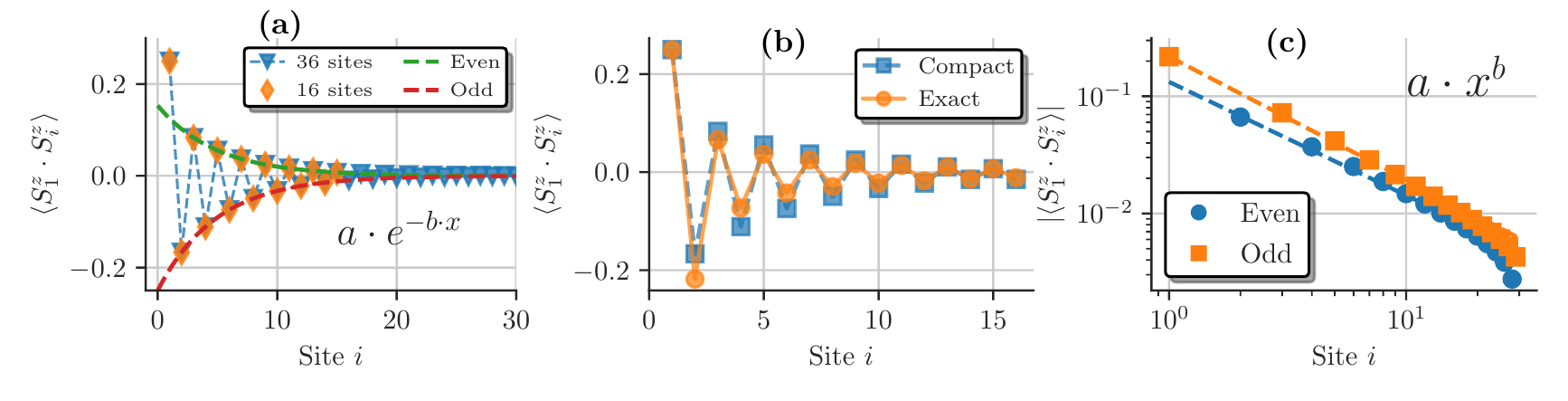}
	\caption{\label{fig:single-csf-corr}(a) Single CSF spin-spin correlation function for the compact ordering vs. chain size and an exponential fit to the even (green dashed line) and odd spin-spin correlation (red dashed line) for the 36-site single CSF in the compact order, with $a_{even}=0.153, b_{even} = 0.203, a_{odd} = -0.25$ and $b_{odd}=0.203$. (b) OBC Single CSF spin-spin correlation function for the compact ordering compared with exact results obtained with the 
		$\mathcal{H}\Phi$ software package\cite{KAWAMURA2017180}
		for 30 sites. (c) Power fit to the even and odd spin-spin correlation for the 30-site exact spin-spin correlation function displayed on a double logarithmic scale, with $a_{even} = 0.133, b_{even} = -0.969, a_{odd} = -0.218$ and $b_{odd} = -1.049$.
	}
\end{figure*}

We extended this study to larger lattices, as shown in Fig.~\ref{fig:single-csf-corr}a, where we find that
for odd- and even lattice spacings the spin-correlation function of the compact reference CSF is exactly
described by an exponential fit  $\braket{\hat S_1^z \cdot S_x^z} = a\cdot \e^{-b x}$ for $x$ even/odd (green and red dashed lines in Fig.~\ref{fig:single-csf-corr}a).
The values of the fit are given by $a_{even} = 0.153$ and  $b_{even} = 0.203$ for even (excluding the first $\braket{S_1^z \cdot S_1^z}$ data point) and $a_{odd} = -1/4$ and $b_{odd} = 0.203$ for odd lattice spacings.
The exact and compact-reference-CSF spin-spin correlation functions for a 30-site lattice with OBC are shown in Fig.~\ref{fig:single-csf-corr}b, which shows that the single CSF results mimic the exact result even for large lattice sizes.
However, the short-range behavior of the exact spin-spin correlation function follows a power law decay\cite{Sandvik2010}, $\braket{\hat S_1^z \cdot \hat S_x^z} \sim a \cdot x^b$, with $a_e = 0.133$ and $b_e = -0.969$ for even and $a_o = -0.218$ and $b_o = -1.049$ for odd lattice spacings in the 30-site case, see Fig.~\ref{fig:single-csf-corr}c.

\begin{figure}
		\includegraphics[width=0.4\textwidth]{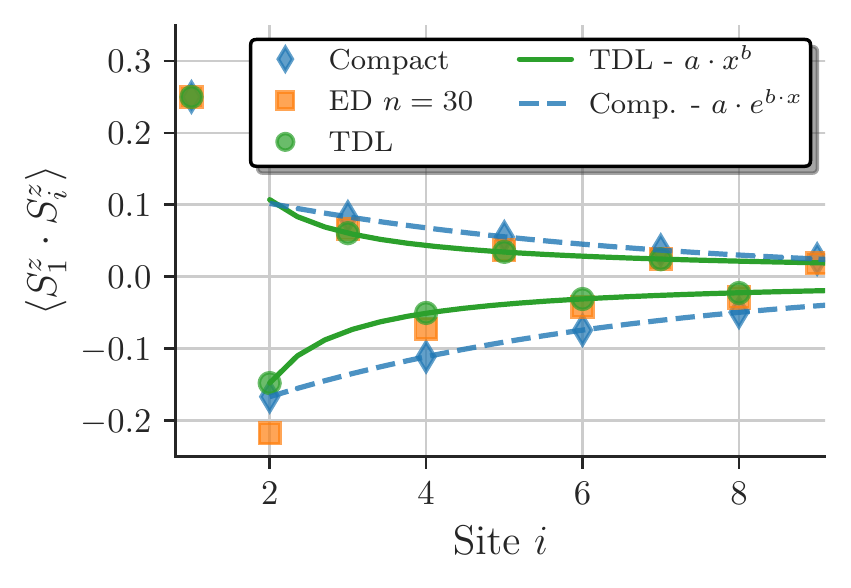}	
	\caption{\label{fig:tdl}Analytic TDL spin-spin correlation function\cite{Sato2005} with power law fits (solid lines) for even and odd lattice sites, with $a_{even} = 0.107,\, b_{even} = -0.820,\, a_{odd} = -0.148$ and $b_{odd} = -0.975$, 
		compact single CSF spin-spin correlation function with exponential fit (dashed lines)
		and $n=30$ exact results (ED) with OBC\cite{KAWAMURA2017180}.}
\end{figure}

Using the method of \emph{generating functions}, Sato \etal\cite{Sato2005} found the exact thermodynamic limit results for $\braket{S_1^z \cdot S_r^z}$ up to $r=7$, which are shown in Table~\ref{tab:exact-spin-corr} along the single CSF and exact diagonalization (ED) results of 30 sites with OBC.
\begin{table}
	\caption{\label{tab:exact-spin-corr}Exact thermodynamic limit spin-spin correlation functions, $\braket{S_j^z S_{j+k}^z}$\cite{Sato2005}, 30-site exact diagonalization (ED) and single CSF results.}
	\renewcommand{\arraystretch}{1.1}
	\begin{tabular}{cccc}
		\toprule
		$k$ & Compact CSF & Exact OBC $n=30$ & Exact TDL\cite{Sato2005} \\
		\midrule
		1 &  -0.166667 		& -0.2174740 		& -0.1477157 		\\ 
		2 &  \p{-}0.083333 	&  \p{-}0.0664877 	& \p{-}0.0606798  	\\ 
		3 & -0.111111 		& -0.0722632 		& -0.0502486 		\\ 
		4 &  \p{-}0.055556 	&  \p{-}0.0370753 	& \p{-}0.0346528 	\\ 
		5 &  -0.074074 		& -0.0417185 		& -0.0308904		\\ 
		6 & \p{-}0.037037 	&  \p{-}0.0251398 	& \p{-}0.0244467 	\\ 
		7 &  -0.049383 		& -0.0286345 		& -0.0224982 		\\ 
		\bottomrule
	\end{tabular}
\end{table}
Based on the entanglement perturbation theory (EPT) approach, Wang and Chung\cite{Wang2012}, derived a relation
of $A(r) = -0.1473 r^{-0.9604}$ with an error of $0.1\%$ for the odd separations, which also fits the even sites with an opposite sign.
Based on a bosonization approach, Hikihara and Furusaki\cite{Furusaki2004} find a critial exponent $b = -1$.
We fitted a power law behavior to the available TDL data\cite{Sato2005}, see Fig.~\ref{fig:tdl},
and gathered the resulting parameters $a$ and $b$ in Table~\ref{tab:corr-length}, along with the
data obtained based on the compact single CSF and 30-site ED results with OBC.
Even though the spin-spin correlation function based on the leading CSF of the compact ordering is exponentially decaying,
as shown in Fig.~\ref{fig:single-csf-corr}a, the short-range behavior, $\braket{\hat S_j^z \cdot \hat S_{j+1}^z}$ -- especially for even lattice spacings -- is quite close to the exact TDL data, as shown in Fig.~\ref{fig:tdl}.

For this reason we also fitted the spin-spin correlation function obtained from the compact-order leading CSF with a power law, $a \cdot x^b$, and gathered the results in Table~\ref{tab:corr-length}.
The number of sites in Table~\ref{tab:corr-length} indicate how many data points of $\braket{\hat S_i^z \cdot \hat S_{i+j}^z}$ were taken into
account for the fitting.
Using only the first 18 lattices yields a critical exponent $b_{even} = -0.979$ very close to the EPT result by
Wang and Chung.
Increasing the considered number of sites to 96, the critical exponent for even lattice separations overestimates the corresponding reference results, but $b_{odd}$ comes closer.
The fact that the spin-spin correlation function obtained by a single CSF is so close to
exact many-body results is striking.

\begin{table}
	\caption{\label{tab:corr-length}Fit of $\braket{S_0^z \cdot S_r^z}$ to $a \cdot x^{b}$ for odd and even lattice sites.}
	{\small
		\begin{tabular}{cccccc}
			\toprule
			Method & \multicolumn{2}{c}{Compact CSF} & ED OBC & EPT\cite{Wang2012} & TDL\cite{Sato2005} \\
			\midrule
			\# sites & 18 & 96 & 30  & $\infty$ & 7  \\
			\midrule
			$a_{even}$ 	& \p{-}0.174 & \p{-}0.201 	& \p{-}0.133 &  \textendash 	& \p{-}0.107 \\
			$b_{even}$ 	& -0.979 	 &  -1.147 		&  -0.969 	 &  -0.9604 		&   -0.820 \\
			Err[\%] 	&  0.3   	 &   0.3 		& 0.1 		 &  0.1 			&   0.0 	\\
			\midrule
			$a_{odd}$ & -0.178  & -0.183 & -0.218 & -0.1473 & -0.148 \\
			$b_{odd}$ & -0.726  & -0.880 & -1.049 & -0.9604 & -0.975 \\
			Err[\%]   & 0.8 	& 0.7 	 &  0.1	  &   0.1 	&  0.0 	 \\
			\bottomrule
		\end{tabular}
	}
\end{table}

\section{\label{sec:fciqmc}GUGA-FCIQMC calculations}

In this section we study the scaling of the (increased) weight of the leading CSF for larger lattice sizes and 
the effect it has on GUGA-FCIQMC calculations, as the method usually benefits from a \enquote{more single reference and sparse} 
character of the sampled wavefunction. Details on the GUGA-FCIQMC method can be found in Appendix~\ref{sec:guga-fciqmc} and Refs.~\citen{Dobrautz2019, Guther2020} and 
computational details and sample input files can be found in the Supplemental Material\cite{SI}.

\begin{figure}
	\includegraphics[width=0.4\textwidth]{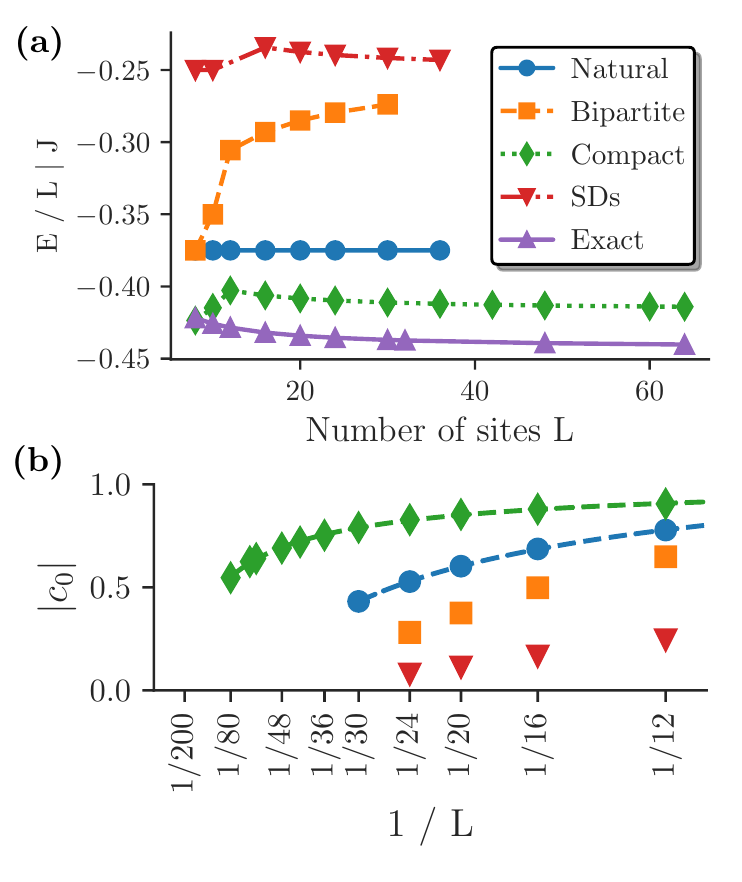}
	\caption{\label{fig:ref-e-obc-chain}(a) Open BC single CSF energies per site of different orderings vs. the 
		number of sites compared with exact Bethe Ansatz\cite{Karbach1998} and ED results, obtained with $\mathcal{H}\Phi$\cite{KAWAMURA2017180}. (b) Open BC weights of the leading CSFs of different orderings vs. the inverse lattice size. 
	}
\end{figure}

As we have identified the single CSF energy as a good indicator for a more optimal ordering, we show it 
for the different orderings in Fig.~\ref{fig:ref-e-obc-chain}a compared to the exact energy of the Heisenberg model, 
obtained with the $\mathcal{H}\Phi$ software package\cite{KAWAMURA2017180},
with OBC as a function of the number of sites. 
It can be seen that the single CSF energy in the compact ordering is closest and in fact almost parallel to the 
exact energy.

\begin{figure}
	\includegraphics[width=0.5\textwidth]{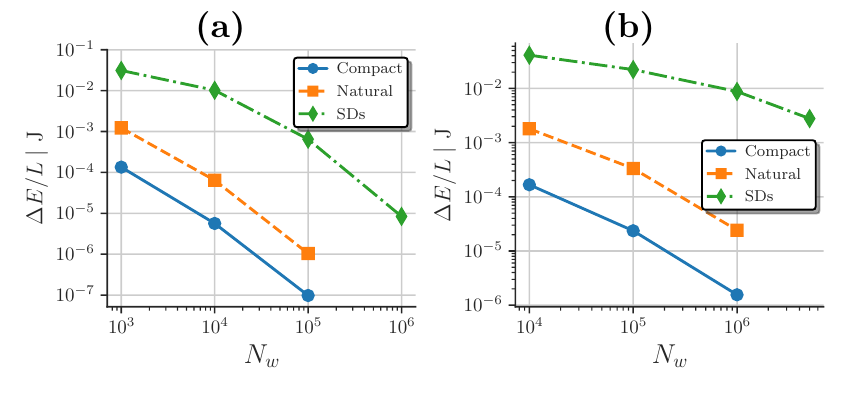}
	\caption{\label{fig:neci-heisenberg}Difference of GUGA-FCIQMC energy per site results compared to ED results\cite{KAWAMURA2017180} of the 20- (a) and 30-site (b)  1D Heisenberg model with open boundary conditions for different orderings (natural and compact) and a SD-based results (SDs) versus the number of walkers $N_w$.}
\end{figure}

In Fig.~\ref{fig:ref-e-obc-chain}b we show the weights of the leading CSF, $\abs{c_0}$, obtained from GUGA-FCIQMC calculations 
for the different orderings as a function of the inverse lattice size, $1/L$, with OBC.
The effect that the weight of the leading CSF in the \emph{compact} ordering is substantially larger compared to the other orderings becomes even more pronounced for larger lattices. 
This increased weight has a very beneficial 
influence on the convergence of GUGA-FCIQMC calculation for finite lattices.
To demonstrate this, Fig.~\ref{fig:neci-heisenberg} shows the energy difference to numerically exact DMRG results, obtained with \texttt{BLOCK}\cite{Chan2002, Chan2008, Sharma2012, OlivaresAmaya2015}, 
for the (a) 20- and (b) 30-site Heisenberg model with OBC for the compact and natural ordering and a SD-based calculations as a function of the number of walkers, $N_w$.

Using the spin-adapted GUGA-FCIQMC calculation with the natural ordering is an order of magnitude more accurate 
for a given number of walkers $N_w$ compared to the standard SD-based implementation. 
In addition, using the compact ordering yields an additional order of magnitude in accuracy.
As we show below, the increased weight of the leading CSF induced by site reordering in the GUGA-scheme is not restricted to the Heisenberg model. 

\subsection{\label{sec:extension}Extension to Hubbard and \emph{ab initio} models.}

Here we extend our study from a pure spin-model to Fermionic problems in form of the Hubbard model and 
to \emph{ab initio} Hamiltonians in form of chains of equally spaced hydrogen atoms.
This entails a much larger Hilbert space size, as the orbitals/sites can also be 
empty or doubly occupied. 
To stay as close as possible to our Heisenberg study above, we choose the parameters of the models in such a way (localized bases, large $U/t$ and hydrogen atom separation) so that the ground states are dominated by states with entirely singly-occupied/open-shell orbitals.

\begin{figure*}
	\includegraphics[width=\textwidth]{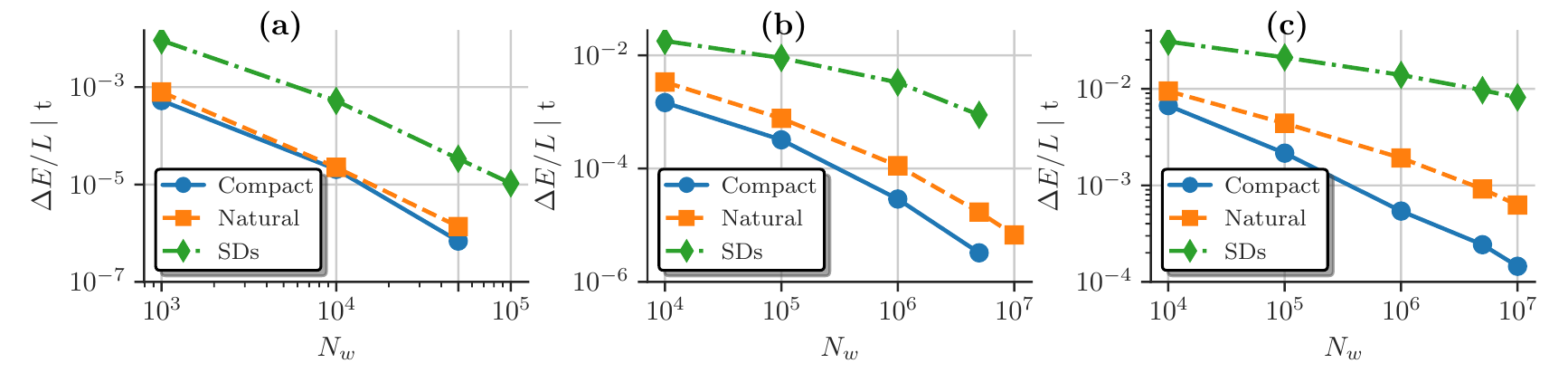}
	\caption{\label{fig:neci-hubbard}
		Difference of GUGA-FCIQMC energy per site results compared to $M=500$ DMRG reference results\cite{Chan2002, Chan2008, Sharma2012, OlivaresAmaya2015} of the 10- (a), 20- (b) and 30-site (c) 1D Hubbard model with open boundary conditions for different orderings (natural and compact) and SD-based results (SDs) versus the number of walkers $N_w$.}
\end{figure*}

Fig.~\ref{fig:neci-hubbard}a-c shows the difference of GUGA-FCIQMC energy per site results compared to numerically 
exact DMRG results\cite{Chan2002, Chan2008, Sharma2012, OlivaresAmaya2015} 
for the 10-, 20- and 30-site Hubbard model with $U/t = 16$ and OBC. 
Similar to the Heisenberg results, using a spin-adapted formulation and the compact ordering yields 
an order of magnitude more accurate results compared to a SD based calculation. 
Albeit not as drastic as for the Heisenberg model, also the weight of the leading CSF, $\abs{c_0}$, is 
substantially increased for the Hubbard model calculations, as shown in Table~\ref{tab:ref-weights-models}.
\begin{table}
	\caption{\label{tab:ref-weights-models}Weight of the leading CSF [\%] for the Heisenberg, Hubbard and Hydrogen chains with OBC for the natural and compact ordering.}
	{\small
		\begin{tabular}{cccccccccc}
			\toprule
			Order &  & \multicolumn{2}{c}{Heisenberg}  & \multicolumn{3}{c}{Hubbard} & \multicolumn{3}{c}{Hydrogen} \\
			\midrule
			\# sites & & 20 & 30 & 10 & 20 & 30 & 10 & 20 & 30 \\
			\midrule
			Compact & & 85.4 & 79.0	& 87.3 & 77.1 & 68.4 	& 74.0 & 54.9 & 46.8 \\
			Natural & & 60.2 & 43.0 & 78.9 & 54.7 & 43.3 	& 67.2 & 41.3 & 36.1 \\
			\bottomrule
		\end{tabular}
	}
\end{table}

\begin{figure*}
	\includegraphics[width=\textwidth]{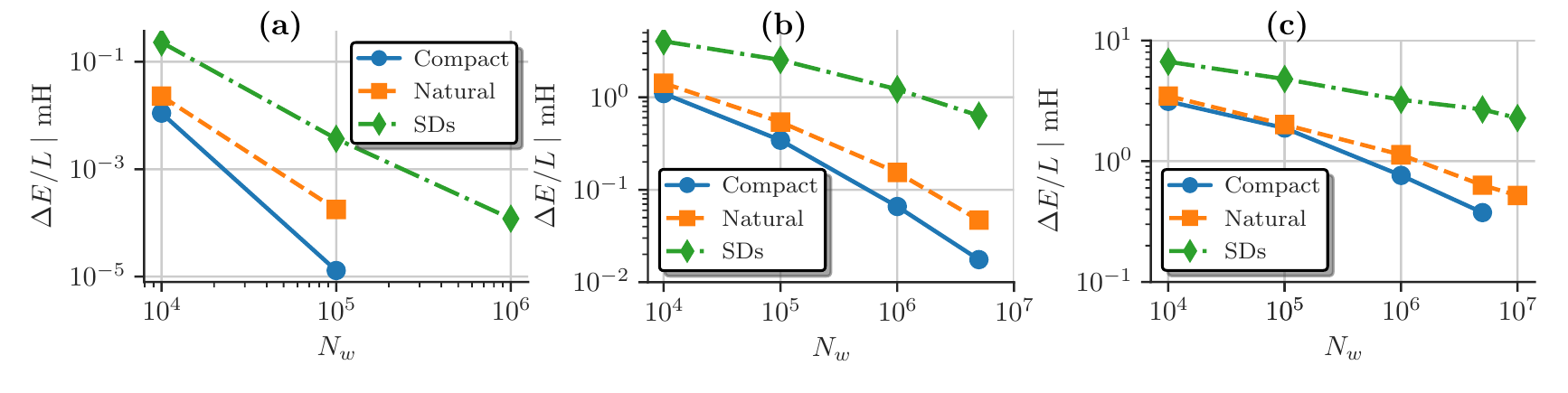}
	\caption{\label{fig:neci-hydrogen}Energy difference per site to the $M=500$ DMRG reference results of the 10- (a), 20- (b) and 30-site (c) hydrogen chain in a minimal basis for different orderings (natural and compact) and SD-based results (SDs) versus the number of walkers $N_w$.}
\end{figure*}

As an example of an \emph{ab initio} model system, 
we study 1D hydrogen chains, recently studied to benchmark various computational physics and chemistry approaches\cite{Motta2017, Motta2020}. 
Fig.~\ref{fig:neci-hydrogen}a-c shows the difference of GUGA-FCIQMC energy per site results compared to numerically exact DMRG 
calculations\cite{Chan2002, Chan2008, Sharma2012, OlivaresAmaya2015} for a 10-, 20- and 30-site hydrogen chain in a STO-3g basis set for the different orderings
and SD-based calculations as a function of the number of walkers. 
The inter-hydrogen separation was 3.6 \AA$\,$ and we used localized orbitals. 
The results are very similar to the Heisenberg and Hubbard model discussed above, where again, 
performing spin-adapted calculations with a compact ordering yields results an order of magnitude more 
accurate for a given number of walkers compared to the standard SD-based FCIQMC method.
In addition the compact ordering increases the weight of the leading CSF compared to the natural ordering, 
as shown in Table~\ref{tab:ref-weights-models}. 

\subsection{\label{sec:dmrg}Comparison with DMRG reordering}

It is well known that orbital reordering is crucial in DMRG.
However, as discussed in our earlier works~\citen{LiManni2020, LiManni2021}
the reordering we seek in the context of spin couplings 
differs from the one in DMRG, both in motivation and in aim. 
In the context of DMRG, site reordering is very important for convergence
with respect to the bond dimension (M),
and relies on concepts of entanglement and quantum (mutual) 
information~\cite{Ali2021, Legeza2003, Moritz2005, Rissler2006, Chan2008, Barcza2011, Mitrushchenkov2011, Keller2014, Chan2002, OlivaresAmaya2015}. 
Our reordering schemes are strictly motivated by the intrinsic mechanisms of the cumulative spin couplings 
and aim at the compression of wave functions expanded in CSFs.
The former is bound to the concept of locality, while our reordering is non-local.

\begin{figure}
	\includegraphics[width=0.4\textwidth]{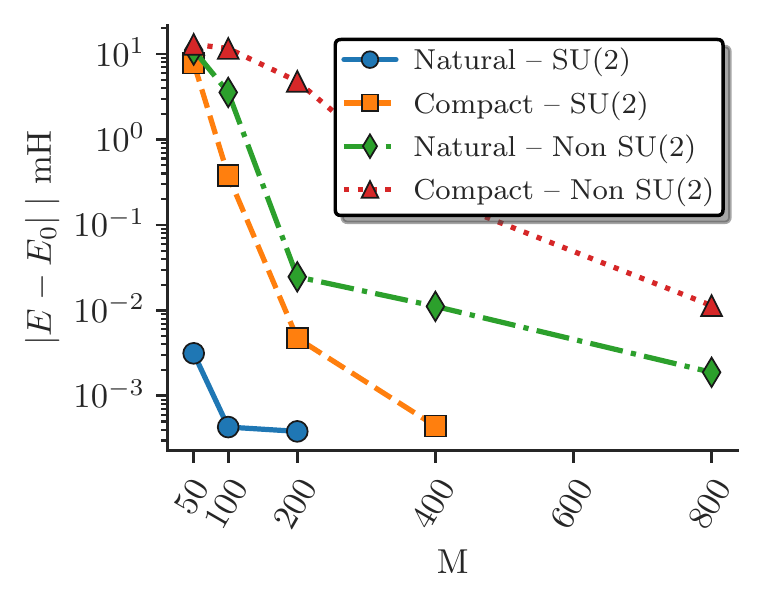}
	\caption{\label{fig:dmrg-hydrogen-chain}30-site hydrogen chain DMRG energy difference to $M=500$ Fiedler (natural order) reference result for different orbital ordering with and without spin-adaptation versus matrix dimension $M$.}
\end{figure}

In this section we provide a numerical proof that the best reordering in GUGA 
is not necessarily the best in DMRG, by analyzing the DMRG convergence using the natural and the 
optimal ordering for a chain of 30 hydrogen atoms.
We used the \texttt{BLOCK} DMRG code\cite{Chan2002, Chan2008, Sharma2012, OlivaresAmaya2015}, which
is able to use $\mathrm{SU}(2)$ symmetry and allows user-defined orbital orderings.
We used the standard Fiedler algorithm~\cite{Barcza2011, Fiedler1973, Fiedler1975, Juvan1992} 
to find the optimal order for the DMRG calculation, which  yielded the natural order as a result. 
(Computational details and sample input files can be found in the Supplemental Material\cite{SI})
We then compared results obtained with the natural/Fiedler ordering
and our \enquote{optimal} compact ordering with and without $\mathrm{SU}(2)$ conservation.
As a reference we used a well-converged $M=400$ result using $\mathrm{SU}(2)$ and natural ordering.
Fig.~\ref{fig:dmrg-hydrogen-chain} shows the results of this study.

A one-dimensional hydrogen chain, with a large atom separation of 3.6\AA, is an optimal case for a MPS based algorithm like DMRG.
This is reflected in the incredibly fast convergence of the natural ordered $\mathrm{SU}(2)$ results with a matrix dimension of $M=100$.
Similar to the GUGA-FCIQMC results, making use of the inherent $\mathrm{SU}(2)$ symmetry is very beneficial for the DMRG convergence. 
However, the 30-site hydrogen results indeed show a difference between the natural and the compact ordering scheme.
While the natural order results with $\mathrm{SU}(2)$ symmetry (blue circles) are already converged for $M=100$, the compact ordering scheme, 
which introduces some long range interactions, requires a matrix dimension of $M=400$ to converge to similar levels of accuracy ($<10^{-2}$ mH).

This demonstrates that the optimal ordering scheme for the GUGA framework differs from the DMRG one. 
While the latter is based on locality and entanglement arguments,
the cumulative spin-coupling in the GUGA enables the inclusion of renormalization-group concepts to
render the description of strongly-correlated many-body systems more compact (a non-local concept).

\section{\label{sec:conclusion}Conclusion}

In this work we demonstrate a novel combined 
symmetric and unitary group approach, applied to the one-dimensional spin-$\frac{1}{2}$ 
Heisenberg model, which yields a more compact ground state wave function.
We find that a specific ordering of the underlying lattice sites, governed
by the symmetric group, $\mathcal{S}_n$, combined with the cumulative
spin-coupling of the unitary group approach, $\mathrm{U}(n)$, resembles
a block-spin/real-space renormalization group.
This induces a more \emph{compact} description of the ground state 
where the most important CSF has a much higher weight than in the natural order.
We derive an analytic formula for this compact CSF for the 1D Heisenberg model,
and find a general description of this \emph{compact} ordering which is easily applicable to 1D lattices of arbitrary size.
We find that this state, up to leading order, already captures with high accuracy 
the spin-spin correlation behavior of the exact ground state wave function. 
A more compact ground state facilitates spin-adapted GUGA-FCIQMC calculations for larger lattice sites.
We compare this found compact ordering to optimal
ordering for DMRG calculations based on quantum mutual information, and find that they differ. 
Finally, we show that this concept also applies for more general lattice models, like the Hubbard model, 
and even to \emph{ab initio} quantum chemical systems, in form of one-dimensional hydrogen chains.
In future work we will investigate the utility of this combined spin-adapted unitary and symmetric group approach in more complex systems, including frustrated spin systems. 

\section*{acknowledgments}
The authors gratefully acknowledge the support of the Max Planck Society.

\appendix

\section*{Appendix}

\section{\label{sec:spin-free-heisenberg}The spin-free formulation of the Heisenberg model}

Expressing the local spin operators as\cite{Paldus2012}
\begin{equation}\label{eq:local-spin-op}
\hat S_i^k = \frac{1}{2} \sum_{\mu,\nu = \uparrow,\downarrow} \sigma_{\mu,\nu}^{k} a_{i,\mu}^\dagger a_{i\nu},
\end{equation}
with the Pauli matrices\cite{Pauli1925}
\begin{equation}\label{eq:pauli-matrices}
\sigma^x = 
\begin{pmatrix}
0	&	\phantom{-}1 \\	1 & \phantom{-}0 \\
\end{pmatrix}, \quad 
\sigma^y = 
\begin{pmatrix}
0 & -i \\ i & \phantom{-}0 \\
\end{pmatrix}, \quad 
\sigma^z = 
\begin{pmatrix}
1 & \phantom{-}0 \\ 0 & -1 \\
\end{pmatrix}
\end{equation}
and the fermionic creation (annihilation) operators, $a_{i,\mu}^{(\dagger)}$ of electrons with spin $\mu$ in spatial orbital $i$. 
This results in the explicit expressions
\begin{align}
\hat S_i^x &= \frac{1}{2}\left( a_{i\u}^\dagger a_{i\d} + a_{i\d}^\dagger a_{i\u} \right), \\
\hat S_i^y &= \frac{i}{2}\left( a_{i\d}^\dagger a_{i\u} - a_{i\u}^\dagger a_{i\d} \right), \\
\hat S_i^z &= \frac{1}{2}\left( n_{i\u} - n_{i\d} \right),
\end{align}
where $n_{i\mu} = a_{i\mu}^\dagger a_{i\mu}$ is the fermionic number operator of orbital $i$ and spin $\mu$.
If we express $\hat{\mathbf{S}}_i \cdot \hat{\mathbf{S}}_j$ as 
\begin{equation}\label{eq:spin-corr-start}
\hat{\mathbf{S}}_i \cdot \hat{\mathbf{S}}_j = \hat{{S}}_i^z \cdot \hat{{S}}_j^z + \hat{{S}}_i^x \cdot \hat{{S}}_j^x + \hat{{S}}_i^y \cdot \hat{{S}}_j^y
\end{equation}
and consequently the individual terms as 
\begin{align}
\hat{{S}}_i^z \cdot \hat{{S}}_j^z =& \frac{1}{4}\left(n_{i\u} - n_{i\d}  \right)\left(n_{j\u} - n_{j\d}  \right), \\
\hat{{S}}_i^x \cdot \hat{{S}}_j^x =& \frac{1}{4}
\left( a_{i\u}^\dagger a_{i\d} + a_{i\d}^\dagger a_{i\u} \right) 
\left(a_{j\u}^\dagger a_{j\d} + a_{j\u}^\dagger a_{j\d} \right) \nonumber\\
=& \frac{1}{4}\Big( \cre{i\u} \ann{i\d} \cre{j\u} \ann{j\d}  + \cre{i\u} \ann{i\d} \cre{j\d} \ann{j\u} \nonumber \\
&+ \cre{i\d} \ann{i\u}  \cre{j\u} \ann{j\d} + \cre{i\d} \ann{i\u} \cre{j\d} \ann{j\u} \Big), \\
\hat{{S}}_i^y \cdot \hat{{S}}_j^y =& -\frac{1}{4}\left( \cre{i\d}\ann{i\u} - \cre{i\u} \ann{i\d}\right) 
\left( \cre{j\d}\ann{j\u} - \cre{j\u}\ann{j\d}\right) \nonumber \\
=& \frac{1}{4}\Big(-\cre{i\d} \ann{i\u} \cre{j\d} \ann{j\u} + 
\cre{i\d} \ann{i\u} \cre{j\u} \ann{j\d} \nonumber \\
&+ \cre{i\u} \ann{i\d} \cre{j\d} \ann{j\u} - 
\cre{i\u} \ann{i\d} \cre{j\u} \ann{j\d}  \Big),
\end{align}  

we can combine the $x$ and $y$ terms as
\begin{align}\label{eq:x-y-sum}
\hat S_i^x \cdot \hat S_j^x + \hat S_i^y \cdot \hat S_j^y =& 
\frac{1}{2}\left( \cre{i\u} \ann{i\d} \cre{j\d} \ann{j\u} + \cre{i\d} \ann{i\u} \cre{j\u}\ann{j\d} \right) \nonumber \\
=& \frac{1}{2}\sum_{\s} \cre{i\s} \ann{i\bar{\s}} \cre{j\bar{\s}} \ann{j\s}.
\end{align}

\underline{For $i\neq j$} we can transform Eq.\eqnref{eq:x-y-sum} to
\begin{equation}
\hat S_i^x \cdot \hat S_j^x + \hat S_i^y \cdot \hat S_j^y =  -\frac{1}{2}\sum_\s \cre{i\s} \ann{j\s} \cre{j\bar{\s}} \ann{i\bar{\s}} = A_{ij}.
\end{equation}
With the spin-free excitation operators, $\hat E_{ij} = \sum_{\s = \u,\d} \cre{i\s}\ann{j\s}$ and $i\neq j$ we can observe
\begin{align}
\hat E_{ij} \hat E_{ji} =& \left(\sum_\s \cre{i\s}\ann{j\s} \right) \left(\sum_\tau \cre{j\tau}\ann{i\tau} \right) \nonumber \\
=&\underbrace{\sum_\s \cre{i\s}\ann{j\s}\cre{j\bar{\s}}\ann{i\bar{\s}}}_{-2A_{ij}} + 
\underbrace{\sum_{\s} \cre{i\s}\ann{j\s} \cre{j\s}\ann{i\s}}_{\num{i\s}(1 - \num{j\s})} \nonumber \\
=& -2 A_{ij} + \sum_\s \num{i\s} - \sum_\s \num{i\s}\num{j\s} \nonumber\\
=& -2A_{ij} + \hat E_{ii} - \sum_\s \num{i\s}\num{j\s},
\end{align}
leading to the relation
\begin{align}
\label{eq:a-to-e}
A_{ij} =& -\frac{1}{2}\left( \hat E_{ij}\hat E_{ji} - E_{ii} + \sum_\s \num{i\s}\num{j\s}\right) \nonumber\\ 
=& -\frac{1}{2} \left(\hat e_{ij,ji} + \sum_\s \num{i\s}\num{j\s}\right), 
\end{align}
where $\hat e_{ij,ji} = \hat E_{ij} \hat E_{ji} - \delta_{jj} \hat E_{ii}$.
With Eq.\eqnref{eq:a-to-e} we can express the spin-spin interaction, Eq.\eqnref{eq:spin-corr-start}, as
\begin{equation}\label{eq:spin-corre-inter}
\hat{\mathbf{S}}_i \cdot \hat{\mathbf{S}}_j  = S_i^z \cdot S_j^z - \frac{1}{2} \left( e_{ij,ji} + \sum_\s \num{i\s}\num{j\s}\right).
\end{equation}
To express Eq.\eqref{eq:spin-corre-inter} entirely in spin-free terms we can rewrite
\begin{align}
\hat S_i^z \cdot \hat S_j^z - &\frac{1}{2}\sum_\s \num{i\s}\num{j\s} = \frac{1}{4}(\num{i\u} - \num{i\d})(\num{j\u} - \num{j\d}) \nonumber \\
&-\frac{1}{2}(\num{i\u}\num{j\u} + \num{i\d}\num{j\d}) \nonumber \\
=& \frac{1}{4}(\num{i\u}\num{j\u} - \num{i\u}\num{j\d} - \num{i\d}\num{j\u} + \num{i\d}\num{j\d})  \nonumber \\
&-\frac{1}{2}(\num{i\u}\num{j\u} + \num{i\d}\num{j\d}) \\
=& -\frac{1}{4}\left( \num{i\u}\num{j\u} + \num{i\u}\num{j\d} + \num{i\d}\num{j\u} + \num{i\d}\num{j\d} \right) \\
=& - \frac{\hat e_{ii,jj}}{4},
\end{align}
which allows us to write the spin-spin correlation function entirely in spin-free terms as 
\begin{equation}\label{eq:spin-free-spin-corr}
\hat{\mathbf{S}}_i \cdot \hat{\mathbf{S}}_j  = -\frac{1}{2}\left(\hat e_{ij,ji} + \frac{\hat e_{ii,jj}}{2} \right).
\end{equation}
It is worth noting that the operator $\hat e_{ii,jj}$ is \emph{diagonal}, and for a Heisenberg model, with explicitly singly 
occupied orbitals, it is identical to one.
This leads to the spin-free formulation of the Heisenberg model, (erratum to Ref.~\citen{dobrautz-phd}) 
\begin{align}\label{eq:spin-free-heisenberg}
H =& J \sum_{\braket{i,j}} \hat{\mathbf{S}}_i \cdot \hat{\mathbf{S}}_j 
= - \frac{J}{2} \sum_{\braket{i,j}}\hat  e_{ij,ji} - \frac{J}{4}\sum_{\braket{i,j}} 1 \nonumber \\
=& - \frac{J}{2} \sum_{\braket{i,j}}\hat e_{ij,ji} - \frac{J N_{b}}{4}, 
\end{align}
where $\braket{i,j}$ indicates the summation over nearest neighbors and $N_b$ is the number of bonds.
This enables a straightforward spin-free implementation of the Heisenberg model in GUGA-FCIQMC, where only 
exchange-type excitations $\hat e_{ij,ji}$ have to be considered.

\section{\label{app:dirac}Dirac identity}

There is an alternative way to derive the spin-free Heisenberg Hamiltonian in terms unitary group generators, $\hat E_{ij}$, based on the Dirac identity\cite{Flocke2002, Dirac1929}.
The Dirac identity\cite{Dirac1929} is given by
\begin{equation}
\label{eq:dirac-id}
(i,j) = \frac{1}{2} + 2\, \Svec{i} \cdot \Svec{j} \quad \rightarrow \Svec{i} \cdot \Svec{j} = \frac{1}{2}\left[(i,j) - \frac{1}{2}\right],
\end{equation}
where $(i,j)$ indicates an exchange of electrons or spins $i$ and $j$.
Eq.~\ref{eq:dirac-id} allowed Flocke and Karwowski\cite{Karwowski1997} to straightforwardly express the 
Heisenberg Hamiltonian, Eq.~\ref{eq:heisenberg-general}, in terms of transpositions of spins $(i,j)$, as 
\begin{equation}
\label{eq:heisenberg-sga}
\hat H = \frac{1}{2}\sum_{ij} J_{ij} (i,j) - \frac{1}{4}\sum_{ij} J_{ij}.
\end{equation}
If we now use the observation by Flocke\cite{Flocke2002} that the action of the operator $(\hat E_{ij}\hat E_{ji} - \mathds{1})$ on a \enquote{Heisenberg wavefunction} (exclusively singly occupied orbitals /spins, no empty sites) $\ket{\phi}$ is identical to a transposition
\begin{equation}
\label{eq:flocke-id}
(\hat E_{ij} \hat E_{ji} - \mathds{1}) \ket{\phi} = (i,j)\ket{\phi},
\end{equation}
and similarly that the action of $\hat E_{ii}$ on $\ket{\phi}$ is just the identity, $\hat E_{ii} \ket{\phi} = \ket{\phi}$, 
we can relate $\hat e_{ij,ji} = (i,j)$ and 
obtain the same Hamiltonian in terms of unitary group generators as in Eq.~\ref{eq:heisenberg-uga}. 

\section{\label{sec:annealing}Simulated annealing}

\begin{figure}
	\includegraphics[width=0.5\textwidth]{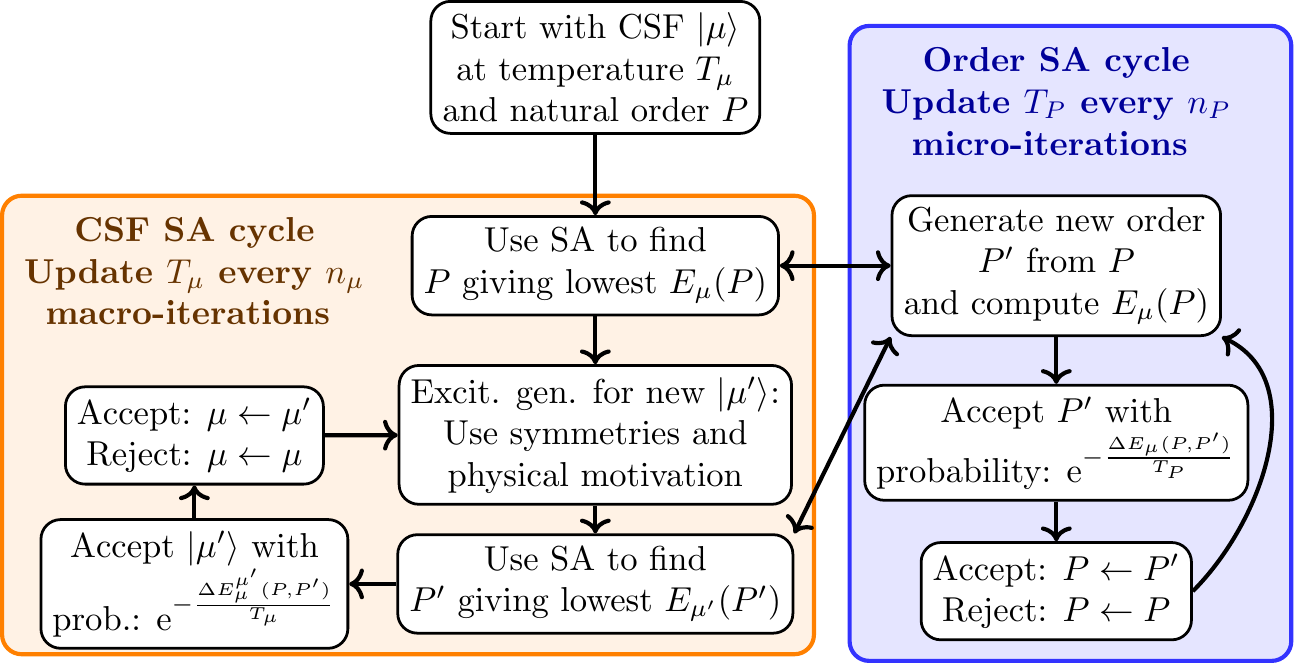}
	\caption{\label{fig:sketch-mix}Sketch of the mixed stochastic minimization procedure.}
\end{figure}

Fig.~\ref{fig:sketch-mix} of our mixed stochastic simulated annealing (SA) approach. 
We start with an arbitrary CSF $\mu$ and order $P$, usually taken as the natural order and the corresponding highest weighted state with alternating $u$'s and $d$'s, $\ket{\mu} = \ket{udududu\dots}$, and temperature $T_\mu$ and $T_P$.
We then use an \emph{inner} SA loop (blue box in Fig.~\ref{fig:sketch-mix}), which finds the optimal order for this CSF $\ket\mu$. 
This is done by suggesting a new order $P' \neq P$, with 2-\cite{Croes1958, Flood1956} and 3-opt modifications\cite{Lin1965} based on the Lin-Kernigham heuristic\cite{Lin1973}. 
In the SA-spirit, this new order $P'$ is always accepted if the diagonal matrix element, $E_\mu(P') = \braopket{\mu}{\hat H(P')}{\mu}$ (Eq.\eqnref{eq:diagonal-exchange}), is lower than $E(P)$, or with probability $p(P'|P) = \e^{-\frac{\Delta E_\mu(P,P')}{T_P}}$, with $\Delta E_\mu(P,P') = E_\mu(P) - E_\mu(P')$. 
We loop over this process of suggesting new orderings $P'$ and lower the temperature $T_P$ every $n_P$  micro-iterations by 
a user defined ratio. This is done until convergence or a set amount of micro-iterations is reached. 

The final ordering $P$ is then fed-back to the main SA cycle (orange box in \ref{fig:sketch-mix}), where 
a new CSF $\ket{\mu'}$ is suggested stochastically. For this we use the \emph{excitation generation} routines in the 
GUGA-FCIQMC method\cite{Dobrautz2019, Guther2020}. 
And for this new CSF $\ket{\mu'}$ we again use the inner loop (blue) to find the optimal ordering $P'$ yielding the lowest 
$E_{\mu'}(P')$. The new state $\ket{\mu'}$ is then accepted in the macro-cycle, if its energy, 
$E_{\mu'}(P') = \braopket{\mu'}{\hat H(P')}{\mu'}$, is lower than $E_\mu(P)$ or with probability 
$p(\mu',P'|\mu, P) = \e^{-\frac{\Delta E_{\mu}^{\mu'}(P,P')}{T_\mu}}$, with 
$\Delta E_{\mu}^{\mu'}(P,P') = E_{\mu}(P) - E_{\mu'}(P')$. 
We also loop over this macro-cycle, where the temperature $T_\mu$ is lowered every $n_\mu$ macro-iterations by 
a user defined ratio, and do this until convergence or a set amount of macro-iterations is reached. 

For small enough systems ($<30$ sites), where the Hilbert space is not yet too large, we can skip the stochastic part (orange box in \ref{fig:sketch-mix}) and perform the SA to find the optimal order (blue box in Fig.~\ref{fig:sketch-mix}) for every state 
in the Hilbert space, as this is an embarrassingly parallel task.

\section{\label{sec:guga-fciqmc}The GUGA-FCIQMC method}

Here, we concisely summarize the main concepts of the GUGA-FCIQMC method. 
An in-depth description of the algorithm can be found in the literature~\cite{dobrautz-phd, Dobrautz2019}.
The FCIQMC algorithm~\cite{Alavi2009, Alavi2010} is based on the 
imaginary-time ($\tau = \text{i}t$) version of the  Schr{\"o}dinger equation,
\begin{equation}
\small
\label{eq:imag-schrodinger}
\frac{\der \ket{\Psi(\tau)}}{\der \tau} = - \hat H \ket{\Psi(\tau)} \; \stackrel{\int \text{d} \tau }{\rightarrow} \; \ket{\Psi(\tau)} = \e^{- \tau \hat H}\ket{\Phi(0)}.
\end{equation}
Integrating Eq.~\eqref{eq:imag-schrodinger} and performing a first-order Taylor expansion yields 
an iterable expression for the eigenstate, $\ket{\Psi(\tau)}$
\begin{equation}
\label{eq:first-order-approx}
\Psi(\tau+\Delta \tau) \approx \left(1 - \Delta \tau \hat H \right) \Psi(\tau),
\end{equation}
which yields the ground state, $\ket{\Psi_0}$, in the long time limit, $\tau \ra \infty$.
In FCIQMC, the full wave function, $\ket{\Psi(\tau)}$, is sampled stochastically by a set of so-called \emph{walkers} and yields estimates for the
ground- and excited-state~\cite{Blunt2015-excited} energies and properties~\cite{Blunt2017} 
via the one- and two-body reduced density matrices.~\cite{Alavi2014}
More details can be found in the literature,~\cite{Alavi2009, Alavi2010} especially in the recently published review article~\citen{Guther2020} and references therein.

In its original implementation FCIQMC is formulated in $m_s$-conserving SDs, 
and thus, does not conserve the total spin quantum number, $S$. 
For interpretability, control, and improved convergence properties,
it is useful to impose the $SU(2)$ spin-rotational symmetry. 
We recently developed a spin-adapted implementation of FCIQMC in our lab\cite{Dobrautz2019, dobrautz-phd, Dobrautz2021}, based on the UGA\cite{Paldus1974, Paldus1975, Paldus1976, Paldus1977, Biedenharn1963, Brooks1979, Brooks1980, DownwardRobb1977, Matsen1974} and its graphical extension (GUGA).~\cite{Shavitt1977,Shavitt1978}.
It was originally conceived for \emph{ab initio} quantum chemical systems, but we recently applied it to study lattice models, like the two-dimensional Hubbard model\cite{Guther2018, Dobrautz2019b, yun2021benchmark}.

The UGA is based on the spin-free formulation of 
quantum chemistry~\cite{Matsen1964} and was pioneered by Paldus\cite{Paldus1974, Paldus2020}, 
who found an efficient usage of the Gel'fand-Tsetlin basis--a general basis for any unitary group $U(n)$\cite{gelfand-1, gelfand-2, gelfand-3}--to the electronic structure problem\cite{Paldus1974, Paldus1976, Paldus1977}.
Based on this work, Shavitt developed the graphical unitary group approach (GUGA),~\cite{Shavitt1977, Shavitt1978, Shavitt1981} which provides 
an elegant and highly effective way to calculate Hamiltonian matrix elements 
between these spin-adapted basis states, also called configuration state functions (CSFs).

%

%

\end{document}


\author{Werner Dobrautz}
\email{dobrautz@chalmers.se}
\affiliation{%
	Max Planck Institute for Solid State Research, Heisenbergstr. 1, 70569 Stuttgart, Germany
}%
\affiliation{
	Department of Chemistry and Chemical Engineering, 
	Chalmers University of Technology, 41296 Gothenburg, Sweden
}

\author{Vamshi M. Katukuri}
\affiliation{%
	Max Planck Institute for Solid State Research, Heisenbergstr. 1, 70569 Stuttgart, Germany
}%

\author{Nikolay A. Bogdanov}
\affiliation{%
	Max Planck Institute for Solid State Research, Heisenbergstr. 1, 70569 Stuttgart, Germany
}%

\author{Daniel Kats}
\affiliation{%
	Max Planck Institute for Solid State Research, Heisenbergstr. 1, 70569 Stuttgart, Germany
}%

\author{Giovanni Li Manni}
\affiliation{%
	Max Planck Institute for Solid State Research, Heisenbergstr. 1, 70569 Stuttgart, Germany
}%

\author{Ali Alavi}
\affiliation{%
	Max Planck Institute for Solid State Research, Heisenbergstr. 1, 70569 Stuttgart, Germany
}%
\affiliation{
	Dept of Chemistry, University of Cambridge, Lensfield Road, Cambridge CB2 1EW, United Kingdom
}%

\title{Supplemental Material: Combined unitary and symmetric group approach applied to low-dimensional spin systems}

\date{\today}

\maketitle

In addition to this document, the Supplemental Material (which can be found 
at~\citen{si}) of this manuscript contains: 
\code{dmrg.conf_10site_heisenberg_obc}, 
\code{dmrg.conf_10site_heisenberg_obc_compact_order}, 
\code{dmrg.conf_10site_heisenberg_obc_nonsu2}, 
\code{dmrg.conf_10site_hubbard_obc}, 
\code{dmrg.conf_10site_hydrogen}
the corresponding integral files: 
\code{FCIDUMP_10site_heisenberg_obc},
\code{FCIDUMP_10site_hubbard_obc}, 
\code{FCIDUMP_10site_hydrogen},
and files for user-defined orderings:
\code{20_site_compact_order}
\code{30_site_compact_order}
\code{10_site_compact_order}.
Sample input for the \texttt{LKH} TSP solver for a 8-site problem 
\code{lkh-8site.inp}
\code{lkh-8site.problem}
and input files for the \texttt{NECI} FCIQMC code:
\code{neci-heisenberg-10-site-obc-bipartite.inp}
\code{neci-heisenberg-10-site-obc-compact.inp}
\code{neci-heisenberg-10-site-obc-natural.inp}
\code{neci-heisenberg-10-site-obc-sds.inp}
\code{neci-heisenberg-30-site-obc-compact.inp}
\code{neci-hubbard-10-site-obc-compact.inp}
\code{neci-hydrogen-10-site-compact.inp}. \\

References to Figures, Tables and Equations from the main document are indicated 
by a \enquote{M-} prefix here.

\tableofcontents 

\section{\label{app:comp-details}Computational details}

The DMRG calculations were performed with the \texttt{BLOCK} DMRG code\cite{Chan2002, Chan2008, Sharma2012, OlivaresAmaya2015} with default settings and matrix dimension of $M = 500$, where not otherwise noted. 
The ED calculations for the Heisenberg model (and spin-spin correlations function extraction) were performed with the \texttt{ALPS} software package\cite{Bauer2011, Wallerberger2018} with standard settings.

The \texttt{LKH}\cite{Helsgaun2000, Helsgaun2009, Tins2018} calculations to verify our simulated annealing implementation
were performed with version 2.0.9, available at \url{http://webhotel4.ruc.dk/~keld/research/LKH/} 
and default settings.
Sample input files to reproduce our approach are provided within this Supplemental Material\cite{si}.
We want to note, that although the TSP, for which LKH was implemented, only has positive entries in the cost matrix, as it indicates distances, the code is also able to solve for negative cost matrix entries, which are given by the $X_{ij}(\mu)$ in
\begin{equation}\label{eq:diagonal-exchange}
\braopket{\mu}{\hat H}{\mu} \sim \sum_{j > i} J_{ij} X_{ij}(\mu), \quad J_{ij} =\begin{cases}
J & \text{for NN},\\
0 & \text{else}.
\end{cases}
\end{equation}
If this would be not possible, this could be 
easily fixed by a shift of the $X_{ij}$ elements by a positive constant. 

The simulated annealing calculations were performed with a starting temperature of $T_{0} = 1$, and for 
usually maximally $10^5$ cycles in total, or after convergence of the diagonal matrix elements, Eq.\eqnref{eq:diagonal-exchange} was achieved for 
a long enough period of time. The temperature was lowered by a factor of $\frac{1}{2}$ every 1000 steps. 

The GUGA-FCIQMC calculations for the Heisenberg lattices up to 36 sites, shown in Fig.~M-8 of the main document, were performed with the original\cite{Alavi2009} 
and numerically exact \emph{non-initiator} implementation. 
Above 36 lattice sites, an initiator threshold of $n_{init} = 3$ was chosen and the 
calculations were checked for convergence up to a maximum walker number of $N_w = 10^8$. 
The GUGA-FCIQMC results shown in Figs.~M-9--M-11 of the main document were 
performed with the initiator approximation\cite{Alavi2010} with $n_{init} = 3$. 
Sample input files for the GUGA-FCIQMC calculations can be found within this Supplemental Material\cite{si}.

\section{\label{app:csf-to-sds}CSFs as linear combination of SDs}

Table~\ref{tab:4-site-csf}
shows the expansion of the 4-site CSFs $\ket{u_1d_2u_3d_4}$ and $\ket{u_1u_3d_2d_4}$ 
and Table~\ref{tab:6-site-csf} the 6-site CSFs $\ket{u_1d_2u_3d_4u_5d_6}$, $\ket{u_1u_3u_5d_2d_4d_6}$ and 
$\ket{u_1u_3d_2u_5d_4d_6}$ in terms of SDs following the transformation of Refs.~\citen{Shavitt1981, Harter1976}, where the subscripts indicate the underlying orbital order, as explained in the 
main document. 
To directly compare the expansion of CSFs into SDs under different orbital orderings, 
we \enquote{map} the resulting SDs to the \emph{natural} ordering. 
As an example the CSF $\ket{u_1 u_3 d_2 d_4}$ has the SD 
$\ket{\u_1 \u_3 \d_2 \d_4}$ in it's linear expansion. 
This SD is then reordered to natural ordering, like $\ket{\u_1 \u_3 \d_2 \d_4} \ra \ket{\u_1 \d_2 \u_3 \d_4}$
in the first column of Tables~\ref{tab:4-site-csf} and \ref{tab:6-site-csf}.
We want to emphasize here again, that this expansion of CSFs into SDs is \textbf{not} necessary 
in the (G)UGA approach, but is \textbf{only} done for demonstrative purposes here.

\begin{table}
	\small
	\caption{\label{tab:4-site-csf}4 Site CSFs to SDs}
	\begin{tabular}{ccc}
		\toprule
		& \multicolumn{2}{c}{Coefficients} \\
		\midrule
		SDs & Natural $\ket{u_1d_2u_3d_4}$ & Bipartite = Compact $\ket{u_1u_3d_2d_4}$ \\
		\midrule
		$\u\, \d\, \u\, \d$ &  \p{-}0.5 & -0.5774 \\
		$\u\, \u\, \d\, \d$ &  \p{-}0.0 & \p{-}0.2887 \\
		$\u\, \d\, \d\, \u$ &  -0.5 	& \p{-}0.2887 \\
		$\d\, \u\, \u\, \d$ &  -0.5 	& \p{-}0.2887 \\
		$\d\, \d\, \u\, \u$ &  \p{-}0.0 & \p{-}0.2887 \\
		$\d\, \u\, \d\, \u$ &  \p{-}0.5 & -0.5774 \\
		\bottomrule
	\end{tabular}
\end{table}

\begin{table*}
	\small
	\caption{\label{tab:6-site-csf}6 Site CSFs to SDs}
	\begin{tabular}{cccc}
		\toprule
		& \multicolumn{3}{c}{Coefficients} \\
		\midrule
		SDs & Natural $\ket{u_1d_2u_3d_4u_5d_6}$ & Bipartite $\ket{u_1u_3u_5d_2d_4d_6}$ & Compact $\ket{u_1u_3d_2u_5d_4d_6}$ \\
		\midrule
		$\u\, \u\, \u\, \d\, \d\, \d\,$ & \p{-}0.0\p{000} & -0.5\p{000} & \p{-}0.0\p{000} \\
		$\u\, \u\, \d\, \u\, \d\, \d\,$ & \p{-}0.0\p{000} & \p{-}0.1667 & \p{-}0.4714\\
		$\u\, \u\, \d\, \d\, \u\, \d\,$ & \p{-}0.0\p{000} & \p{-}0.1667 & -0.2357 \\
		$\u\, \u\, \d\, \d\, \d\, \u\,$ & \p{-}0.0\p{000} & \p{-}0.1667 & -0.2357 \\
		$\u\, \d\, \u\, \u\, \d\, \d\,$ & \p{-}0.0\p{000} & \p{-}0.1667 & -0.2357 \\
		$\u\, \d\, \u\, \d\, \u\, \d\,$ & \p{-}0.3536 & \p{-}0.1667 & \p{-}0.1179\\
		$\u\, \d\, \u\, \d\, \d\, \u\,$ & -0.3536 & \p{-}0.1667 & \p{-}0.1179 \\
		$\u\, \d\, \d\, \u\, \u\, \d\,$ & -0.3536 & -0.1667 & -0.1179 \\
		$\u\, \d\, \d\, \u\, \d\, \u\,$ & \p{-}0.3536 & -0.1667 & -0.1179 \\
		$\u\, \d\, \d\, \d\, \u\, \u\,$ & \p{-}0.0\p{000} & -0.1667 & \p{-}0.2357 \\
		$\d\, \u\, \u\, \u\, \d\, \d\,$ & \p{-}0.0\p{000} & \p{-}0.1667 & -0.2357 \\
		$\d\, \u\, \u\, \d\, \u\, \d\,$ & -0.3536 & \p{-}0.1667 & \p{-}0.1179 \\
		$\d\, \u\, \u\, \d\, \d\, \u\,$ & \p{-}0.3536 & \p{-}0.1667 & \p{-}0.1179 \\
		$\d\, \u\, \d\, \u\, \u\, \d\,$ & \p{-}0.3536 & -0.1667 & -0.1179 \\
		$\d\, \u\, \d\, \u\, \d\, \u\,$ & -0.3536 & -0.1667 & -0.1179 \\
		$\d\, \u\, \d\, \d\, \u\, \u\,$ & \p{-}0.0\p{000} & -0.1667 & \p{-}0.2357 \\
		$\d\, \d\, \u\, \u\, \u\, \d\,$ & \p{-}0.0\p{000} & -0.1667 & \p{-}0.2357 \\
		$\d\, \d\, \u\, \u\, \d\, \u\,$ & \p{-}0.0\p{000} & -0.1667 & \p{-}0.2357 \\
		$\d\, \d\, \u\, \d\, \u\, \u\,$ & \p{-}0.0\p{000} & -0.1667 & -0.4714 \\
		$\d\, \d\, \d\, \u\, \u\, \u\,$ & \p{-}0.0\p{000} & \p{-}0.5\p{000} & \p{-}0.0\p{000} \\
		
		\bottomrule
	\end{tabular}
\end{table*}

\section{\label{app:6-site-matrices}6-site Heisenberg Hamiltonian matrices for different orderings}

\begin{figure*}
	\includegraphics[width=0.7\textwidth]{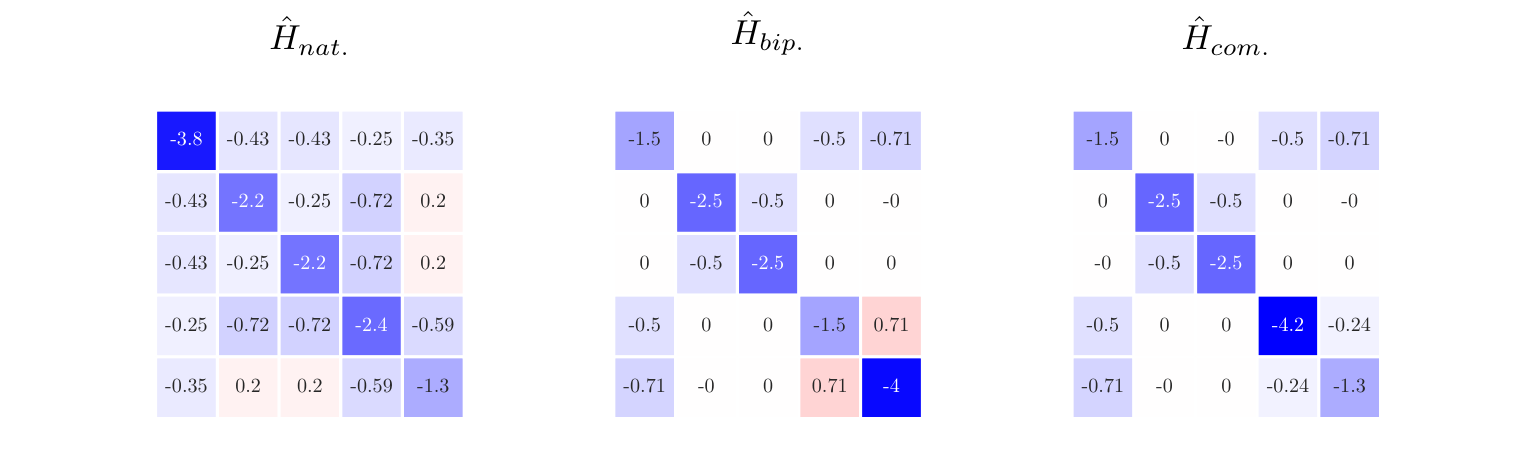}
	\caption{\label{fig:6site-h-pbc}6-site Heisenberg Hamiltonians with PBC in natural (nat.), bipartite (bip.) and compact (com.) order.}
\end{figure*}

\begin{figure*}
	\includegraphics[width=0.7\textwidth]{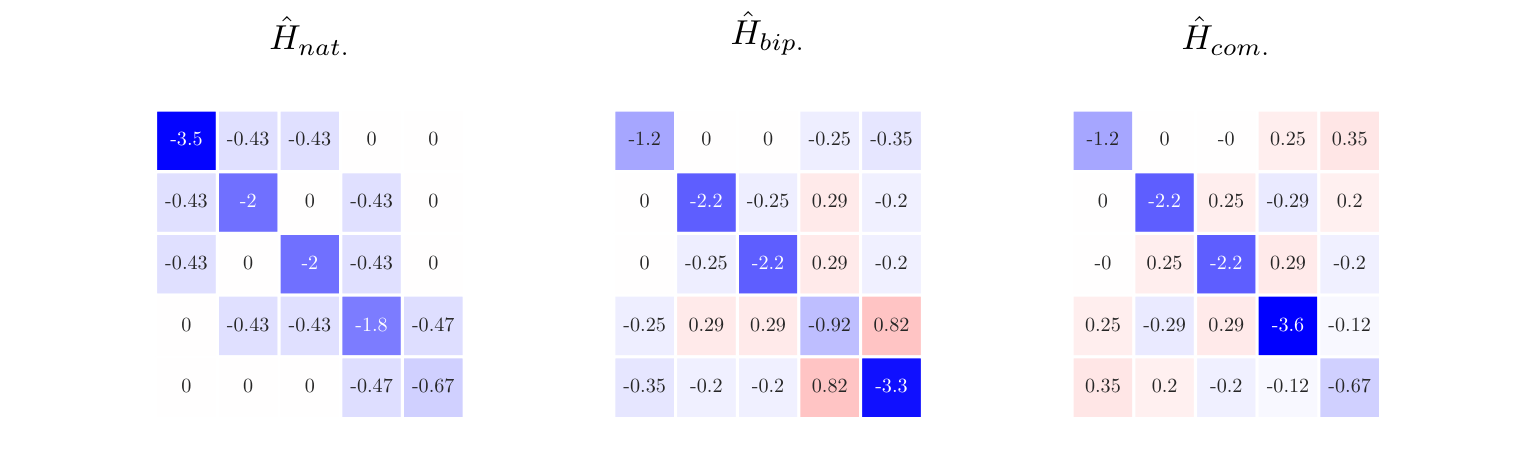}
	\caption{\label{fig:6site-h-obc}6-site Heisenberg Hamiltonians with OBC in natural (nat.), bipartite (bip.) and compact (com.) order.}
\end{figure*}

For demonstrative purposes we want to show the explicit Heisenberg Hamiltonians 
for the 6-site lattice with PBC and OBC in the \emph{natural}, \emph{bipartite} and \emph{compact}
orderings here.
We also want to show the equivalence of the 
cumulative GUGA compact ordering: $((((1 + 3) + 2) + 5) + 4) + 6$, and 
the separate renormalization approach: $((1 + 3) + 2) + ((4 + 6) + 5)$
mentioned in the main text here. 
The Hilbert space in for the 6-site Heisenberg model is given by
\begin{align*}
\ket 1 &= \ket{u d u d u d}, \quad \ket 2 = \ket{u u d d u d}, \quad \ket 3 = \ket{u d u u d d}, \\
\ket 4 &= \ket{u u d u d d}, \quad \ket 5 = \ket{u u u d d d},
\end{align*}
where we omit the ordering subscript for readability.
For PBC the Hamiltonians in natural, bipartite and compact ordering are 
given in Fig.~\ref{fig:6site-h-pbc}
and for OBC in Fig.~\ref{fig:6site-h-obc}.

\section{\label{app:compare-square}Ordering for $4 \times 2$ ladder}

\begin{figure*}
	\includegraphics[width=\textwidth]{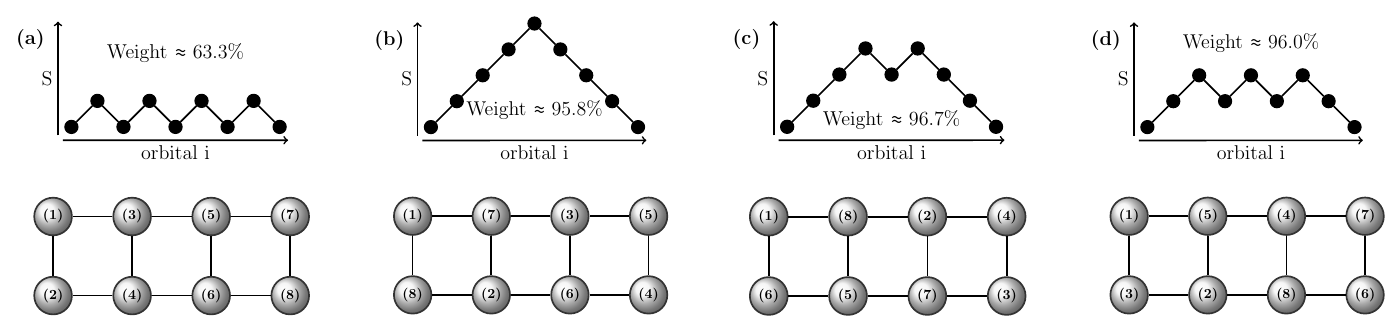}
	\caption{\label{fig:4x2} Weight and reference for the $4\times 2$ ladder with PBC 
		in (a) natural (equivalent to the order A in Flocke and Karwowski\cite{Karwowski1997}), (b) compact 
		and (c) order B in Flocke and Karwowski\cite{Karwowski1997}, as well as (d) compact order with OBC.}
\end{figure*}

Fig.~\ref{fig:4x2} shows different orderings for a $4 \times 2$ Heisenberg ladder with PCB (a-c) and OBC (d). 
We tested the effect of the two different ordering identified in Flocke and Karwowski\cite{Karwowski1997} 
on the reference weight in the UGA. 
The natural order, Fig.~\ref{fig:4x2}a, is the preferred one in Flocke and Karwowski\cite{Karwowski1997} and 
equivalent to our natural order concept, while the order shown in Fig.~\ref{fig:4x2}b was found to be detrimental to Flocke and Karwowskis approach and is equivalent to a bipartite ordering and yields a drastically improved 
reference weight in the UGA. 
Fig.~\ref{fig:4x2}c shows the optimal order and reference CSF for the  $4 \times 2$ Heisenberg ladder with PCB, 
which has an even larger reference weight of $\approx 96.7\%$ compared to the already large weight of $\approx 95.8\%$ in the bipartite order. 
Fig.~\ref{fig:4x2}d shows the optimal order and reference CSF for he  $4 \times 2$ Heisenberg ladder with OCB, 
which has the same structure as the optimal CSF found in the 1D case. 

%

%